\documentclass[%
reprint,
nofootinbib,
amsmath,amssymb,
aps,
showkeys,
onecolumn,
pra,
]{revtex4-2}
\usepackage[utf8]{inputenc} 
\usepackage[T1]{fontenc}    
\usepackage{hyperref}       
\usepackage{url}            
\usepackage{booktabs}       
\usepackage{amsfonts}       
\usepackage{nicefrac}       
\usepackage{microtype}      
\usepackage{xcolor}         
\usepackage{amsthm}
\usepackage[normalem]{ulem} 
\usepackage{array}
\usepackage{amsmath}
\usepackage{amssymb}
\usepackage{graphicx}
\usepackage{mathtools}
\usepackage{footnote}
\usepackage{comment}
\usepackage{cprotect}
\usepackage{accents}
\usepackage{placeins}
\usepackage{cancel}

\usepackage{algorithm}

\usepackage{bbold}
\usepackage{algpseudocode}
\setcitestyle{numbers,square}

\DeclareMathOperator{\R}{\mathbb{R}}

\DeclareMathOperator{\hrho}{\hat{\rho}}

\newcommand{\x}{\underbar{x}}
\newcommand{\y}{\underbar{y}}
\newcommand{\B}{\underbar{B}}

\newcommand{\dd}{\mathop{}\!\mathrm{d}}
\newcommand{\ii}{\mathrm{i}}

\newcommand{\Om}{\bm{\Omega}}
\newcommand{\hOm}{\widehat{\bm{\Omega}}}
\newcommand{\ee}{\mathrm{e}}

\usepackage{dcolumn}
\usepackage{bm}

\usepackage{indentfirst}
\begin{document}

\preprint{APS/123-QED}

\title{Largest eigenvalue and top eigenvector statistics of large Euclidean random matrices} 

\author{Pasquale Casaburi}
\email{pasquale.casaburi@kcl.ac.uk}

\author{Pierpaolo Vivo}

\affiliation{Department of Mathematics, King’s College London, The Strand, London WC2R 2LS, UK}

\begin{abstract}
Euclidean random matrices arise in a wide range of physical systems where interactions are determined by spatial configurations, including disordered media and cooperative phenomena in atomic ensembles. Unlike classical random matrix ensembles, their entries are strongly correlated through the geometry of the underlying random points, making their analytical treatment challenging. While global spectral properties such as the spectral density are relatively well understood, much less is known about extremal eigenvalues and the associated eigenvectors, despite their central role in applications. Here we address the problem of characterising the largest eigenvalue and the corresponding top eigenvector of large Euclidean random matrices with a generic symmetric kernel. For vectors in any dimension $d\geq 1$ drawn independently from a common distribution, we show that both quantities can be computed within a unified replica-based framework. In the large-$N$ limit, the replica saddle-point equations reduce to an eigenvalue problem for a Fredholm integral operator determined by the kernel and by the underlying distribution. We then specialise this general formulation to the quadratic distance kernel, for which the Fredholm problem reduces to a finite set of $d+2$ self-consistent equations. In this case, we obtain an explicit expression for the average largest eigenvalue, fully determined by low-order moments of the underlying distribution, and an analytical characterisation of the density of the top eigenvector's components. We further perform extensive numerical simulations that confirm these predictions. More broadly, our work provides a general framework to access extremal spectral properties of Euclidean random matrices.
\end{abstract}


\maketitle

\section{Introduction}
Random matrix theory (RMT) is a central tool in modern theoretical physics, providing a powerful framework to describe the statistical properties of large complex systems \cite{mehta}. Originally introduced in statistics in the 20s \cite{wishart1928,hsu1939distribution} and later employed in nuclear physics to model the statistics of energy levels of heavy nuclei, it has since found applications across a wide range of fields, including condensed matter physics, quantum chaos, statistical inference, and wireless communications \cite{mehta,PP_libro,JPB_libro}. 
Ensembles of complex hermitian or real symmetric random matrices for which spectral observables can be characterised analytically usually include either (i) rotationally invariant ensembles, whose joint probability density of the real eigenvalues is explicitly known \cite{PP_libro}, or (ii) matrices with independent entries, which can be often analysed using the replica or cavity methods from the physics of disordered systems \cite{Rogers_cavity,PP_cavity_review,PP_cavity_silvia}. 

A very important class of models that lie outside this perimeter is provided by \emph{Euclidean random matrices} (ERMs), whose entries are defined through a deterministic function of the positions of randomly distributed points in an Euclidean space \cite{mezard_parisi_zee}. More precisely, given a set of $N$ random points $\{\x_i\}_{i=1}^N$ in $\mathbb{R}^d$, an ERM is a matrix whose elements take the form

\begin{equation}\label{eq:general_X}
    X_{ij} = f(\x_i,\x_j) \ .
\end{equation}
This class of matrices naturally arises in a variety of physical problems where disorder is encoded in the spatial configuration of interacting elements, rather than in independent random entries \cite{parisi_erm,mezard_parisi_zee,CicutaKrausserMilkusZaccone2018}. Notable examples include vibrational spectra of glasses and amorphous solids \cite{parisi_erm,mezard_parisi_zee}, wave propagation and Anderson localisation in random media \cite{skipetrov_goetschy,ciliberti_anderson}, as well as collective phenomena such as cooperative spontaneous emission in large atomic clouds \cite{skipetrov_goetschy, goetschy_review}. A closely related class of models are distance matrices, where entries depend on pairwise distances between random points and exhibit non-trivial spectral and localisation properties \cite{bogomolny_distance}. 

Despite their broad applicability, the theoretical analysis of ERMs is significantly more challenging than that of ``classical'' random matrix ensembles. The main difficulty stems from the strong correlations between matrix elements, which are induced by the underlying geometry of the random points and are generally difficult to handle analytically \cite{goetschy_review}.  As a consequence, many of the standard techniques of RMT cannot be directly applied. While some analytical progress has been achieved in specific limits or using perturbative and diagrammatic methods such as high-density expansions \cite{Grigera_2011}, mean-field approximations, or mappings to simpler ensembles, most available results are still limited to global spectral properties such as the spectral density \cite{goetschy_review, parisi_erm, bordenave_2,jiang_palle,BattistaMonasson2020}. 

The study of \emph{largest eigenvalue} and \emph{top eigenvector components} for Euclidean matrices has received comparatively much less attention, in spite of being an area of strong relevance for applications, as discussed below. On the one hand, this is somewhat surprising: since the pioneering work by Tracy and Widom \cite{tracy1994_seminal}, extreme eigenvalues of random matrices have been extensively studied for both rotationally invariant ensembles (also in the large deviations regime) \cite{Satya_large_dev} and sparse matrices \cite{PP_susca, kabashima_sparse}. On the other hand, Euclidean matrices are generated by only $\mathcal{O}(N)$ random degrees of freedom, rather than by $\mathcal{O}(N^2)$ independent matrix entries. As a result, their entries are subject to strong geometric constraints inherited from the underlying point configuration. For example, when $f(\x_i,\x_j)$ is the Euclidean distance between points in $\mathbb{R}^d$, the matrix entries must satisfy simultaneous triangle inequalities involving multiple rows and columns. These constraints make the analysis of extreme spectral observables in principle more onerous.

A particularly relevant application, where some analytical control over extreme eigenvalues would be of central importance, is cooperative light-matter interactions in disordered atomic systems. In cold atomic clouds, photon-mediated interactions between randomly distributed atoms lead to collective emission phenomena that cannot be described in terms of independent emitters \cite{skipetrov_goetschy, viggiano_2023, viggiano_2025}. In this setting, the system is described by a matrix whose entries depend on interatomic distances, thus naturally defining a Euclidean random matrix. Its eigenvalues correspond to the decay rates of collective excitation modes, distinguishing between enhanced (superradiant) and suppressed (subradiant) emission processes \cite{viggiano_2023,viggiano_2025}. The extreme eigenvalues are therefore of central importance: the largest correspond to strongly superradiant modes, while the smallest are associated with long-lived subradiant states, both relevant for quantum technologies \cite{viggiano_2023, viggiano_2025, quantum_application}. The associated eigenvectors are physically relevant as well, since their components determine the relative excitation amplitudes of the atoms in each collective decay channel. For instance, the superradiant eigenvector is the fully symmetric excitation shared by all atoms, whereas subradiant modes correspond to orthogonal combinations whose amplitudes interfere destructively and suppress emission \cite{viggiano_2023,viggiano_2025}. Thus, the top eigenvector identifies the spatial profile of the most superradiant decay channel. Characterising the statistics of extreme eigenvalues and their associated eigenvectors is therefore of central importance to understanding the cooperative behaviour of the system.

Despite this direct and important connection to physical problems, a systematic analytical description of extremal eigenpairs in Euclidean random matrices remains largely underdeveloped. Some progress in this direction has been made in \cite{bordenave_1}, where the authors characterise the spectral density of Euclidean random matrices and relate the spectral radius to the Fourier transform of the underlying kernel, although the analysis is restricted to uniformly distributed points in space. Similarly, in \cite{cheng_singer} the spectral density of random kernel matrices is characterised and an upper bound for the spectral norm is derived. However, this bound is not expected to be sharp, and a more precise analysis of the largest eigenvalue statistics remains largely open, with only partial results available under strong regularity assumptions on the kernel function \cite{el_karoui}. The analysis is also restricted to specific ensembles, such as Gaussian vectors or points uniformly distributed on the sphere. A more direct approach to the statistics of extreme eigenvalues is pursued in \cite{chatterjee_huang}, where explicit distributional limits for the largest eigenvalue of kernel matrices are derived. However, this analysis is limited to one-dimensional kernels with uniformly distributed inputs. In physically motivated models, such as those describing cooperative emission in atomic clouds, recent works have addressed the behaviour of the smallest eigenvalues for specific kernels and spatial distributions \cite{viggiano_2025}, revealing non-trivial collective phenomena, but again in restricted settings (e.g., fixed dimension and Gaussian spatial distributions). Overall, while important progress has been made in specific cases, a general analytical framework to characterise extremal eigenvalues of Euclidean random matrices for arbitrary underlying distributions and in any dimension $d\geq 1$ is still missing. This gap is even more pronounced for the statistics of the top eigenvector's components, for which analytical results are virtually absent. 

In this work, we develop a direct analytical approach to characterise the statistics of the top eigenpair of a broad class of Euclidean random matrices. Considering symmetric matrices defined through a generic kernel function $f(\x_i,\x_j)$ of random vectors $\x_i$ in $d$ dimensions, drawn independently from a common distribution P(\x), we employ the \emph{replica method} from the physics of disordered systems to derive closed equations governing the typical value of the largest eigenvalue and the density of top eigenvector's components in the large-$N$ limit. Replica-based methods provide a powerful framework to compute spectral observables in random matrix problems by mapping them to effective statistical mechanics problems \cite{mezard_parisi_zee, parisi_erm}. In particular, the replica method can be employed to access not only global spectral properties, such as densities and resolvents \cite{PP_cavity_review,BattistaMonasson2020}, but also local quantities, including extreme eigenvalues and the associated eigenvectors \cite{Aufiero_2stipendi, PP_susca, Urte,Fyodorov_topology_triv, Doussal_replica}.
For a generic symmetric kernel, the replica saddle-point equations reduce to an eigenvalue problem for a Fredholm integral operator determined by the kernel and by the distribution of the random vectors, providing a compact formulation of the average largest eigenvalue. In the main text, we specialise this formalism to the quadratic distance kernel, for which the Fredholm problem reduces to a finite set of $d+2$ self-consistent equations. In this case, the average largest eigenvalue can be written explicitly in terms of low-order moments of the underlying distribution of points. This provides a transparent framework that applies to a wide range of distributions, including isotropic, independent, and component-wise correlated cases. In the Appendices, we also discuss a centered Gaussian kernel, for which no finite-dimensional reduction of the Fredholm problem is possible, but a closed-form solution is still attainable. In addition, we extend the analysis to the structure of the associated top eigenvector for a generic symmetric kernel. By introducing an auxiliary partition function, we derive an analytical expression for the density of its components, which is determined by the principal eigenfunction of the corresponding Fredholm operator. For the quadratic distance kernel, this reveals a non-trivial geometric structure that reflects the underlying spatial disorder. In particular, we show that the components concentrate on a hypersurface determined by the same parameters controlling the largest eigenvalue, providing a unified description of both eigenvalue and eigenvector statistics. All analytical results have been checked against numerical simulations on randomly generated instances with excellent agreement.

The paper is organised as follows. In Section~\ref{sec:eigenvalue}, we introduce the model and present the replica-based derivation of the largest eigenvalue. We first obtain a general formulation in terms of the Fredholm eigenvalue problem in Eq.~\eqref{eq:Fredholm_problem} for a generic symmetric kernel, and then specialise to the quadratic distance kernel, where the problem reduces to a finite set of self-consistent equations and leads to the explicit expression in Eq.~\eqref{eq:lambda_final_short}. In Section~\ref{sec:eigenvector}, we analyse the statistics of the top eigenvector and derive the density of its components in the large-$N$ limit. We first present the general expression in terms of the Fredholm eigenfunction, given in Eq.~\eqref{eq:T(u)_suitable_for_iid}, and then discuss its explicit form for the quadratic distance kernel in Eq.~\eqref{eq:T_over_Sd-1=0}. In Section~\ref{sec:results}, we focus on the quadratic distance kernel and present analytical solutions for specific classes of distributions, including isotropic and independent cases. We then compare these predictions with extensive numerical simulations, finding excellent agreement. In Section \ref{sec:conclusion}, we discuss the implications of our results and possible extensions. In Appendix~\ref{sec:appendix_A}, we show how the spectral problem for distance matrices admits an exact finite-dimensional reduction involving a set of $(d+2)$ coefficients that fully determine the non-zero eigenvalues and eigenvectors. In Appendix~\ref{sec:appendix_B}, we relax the i.i.d. assumption and show how the replica framework can be extended to correlated disorder of Curie--Weiss type. In Appendix~\ref{sec:appendix_C}, we apply the general Fredholm formulation to a Gaussian kernel, illustrating a case beyond the quadratic distance kernel where no finite-dimensional reduction occurs and the largest eigenvalue and the density of the top eigenvector's components are obtained directly from the integral eigenvalue problem.

\section{Largest eigenvalue}\label{sec:eigenvalue}
Consider $N$ random vectors $\x_i=(x_{i1},...,x_{id})^T\in \R^d$, with joint probability density $P(\x_1,\dots,\x_N)$. Given a symmetric function $f:\R^d\times\R^d \to \R$, we define the real symmetric random $N\times N$ matrix $X$ by

\begin{equation}\label{eq:matrix_X_generic}
    X_{ij}=\frac{f(\x_i,\x_j)}{N}\ , \qquad \forall i,j\in\{1,\dots,N\}\ .
\end{equation}
Using the Courant–Fisher characterisation, the largest eigenvalue of $X$ can be written as 

\begin{equation}\label{eq:courant-fisher}
\lambda_{\mathrm{max}}=
\frac{1}{N}
\max_{\|\vec{v}\|^2 = N}
(\vec{v},X\vec{v})\ ,
\end{equation}
where $(\cdot,\cdot)$ denotes the standard dot product in $\R^N$. With this normalisation and under mild moment assumptions on $P$, the largest eigenvalue is of order one in the large-$N$ limit. Defining an auxiliary partition function at inverse temperature $\beta$

\begin{equation}\label{eq:partition-function}
Z_N(\beta)\coloneq \int \dd \vec{v}~\exp\left\{\frac{\beta}{2} (\vec{v},X\vec{v})\right\} \delta(||\vec{v}||^2-N)\ ,
\end{equation}
in the zero-temperature limit $\beta\to\infty$, the Laplace method together with Eq. \eqref{eq:courant-fisher} gives

\begin{equation}
Z_N(\beta) \underset{\beta \to \infty}{\approx} \exp\left\{\frac{\beta}{2} \max_{\|\vec{v}\|^2 = N}(\vec{v},X\vec{v})\right\} = \exp\left\{\frac{\beta}{2} N \lambda_{\mathrm{max}}\right\}\ ,
\end{equation}
and therefore

\begin{equation}
\lambda_{\mathrm{max}}=\lim_{\beta \to \infty} \frac{2}{\beta N} \log Z_N(\beta)\ .
\end{equation}
To compute the average of $\lambda_{\mathrm{max}}$ we consider the large-$N$ limit and employ the replica trick

\begin{equation}\label{eq:lambda_max_limits}
\langle\lambda_{\mathrm{max}}\rangle=\lim_{N\to \infty} \frac{1}{N}\lim_{\beta \to \infty} \frac{2}{\beta} \lim_{n\to 0} \frac{1}{n} \log \langle Z_N(\beta)^n\rangle\ ,
\end{equation}
where $n$ is initially treated as an integer, and then analytically continued to real values in the vicinity of $n=0$. As usual in replica computations, we assume that the large-$N$ limit can be performed first \cite{noemi,Aufiero_2stipendi,pedestrian}, so that the subsequent zero-temperature and replica limits are taken on the resulting saddle-point expression. The starting point of our analysis is therefore the computation of the averaged replicated partition function \(\langle Z_N(\beta)^n \rangle\), where the average is taken over all possible realisations of the matrix $X$. Since $X$ is fully specified by the random vectors $\{\x_i\}_{i=1}^N$, this corresponds to averaging over their joint distribution $P(\x_1,\dots,\x_N)$, yielding

\begin{equation}
\begin{aligned}\label{eq:Z^n_starting}
    \langle Z_N(\beta)^n \rangle&=\int \prod_{c=1}^{n} [\dd\Vec{v}_c~\delta(||\Vec{v}_c||^2-N)]\int \dd\x_1 \cdots \dd\x_N 
     P(\x_1,\dots,\x_N)~\exp\left\{\frac{\beta}{2N}\sum_{c=1}^{n}\sum_{i,j}^{N} f(\x_i,\x_j) v_{ic}v_{jc}\right\}\ .
\end{aligned}
\end{equation}
We rewrite this average using the Fourier representation of the Dirac delta function

\begin{equation}
\begin{aligned}\label{eq:lambda_introduction}
    \prod_{c=1}^{n} \delta(||\Vec{v}_c||^2-N)=\left(\frac{\beta}{4\pi}\right)^n\int \dd\boldsymbol{\lambda} \exp\left\{-\ii\frac{\beta}{2}\sum_{c=1}^n \sum_{i=1}^N\lambda_c v_{ic}^2 + \ii\frac{\beta}{2}N \sum_{c=1}^n \lambda_c \right\}\ ,
\end{aligned}
\end{equation}
and introduce $n$ densities $\{\rho_c(\x)\}_{c=1}^n$ \cite{noemi}, defined as

\begin{equation}
\begin{aligned}\label{eq:rho_definition}
    \rho_c(\x)\coloneqq \frac{1}{N}\sum_{i=1}^{N} \delta(\x-\x_i) v_{ic}= \frac{1}{N}\sum_{i=1}^{N}\prod_{\ell=1}^d\delta(x_\ell-x_{i\ell})v_{ic}\ .
\end{aligned}
\end{equation}
Using Eq.~\eqref{eq:rho_definition} we can write

\begin{equation}
\begin{aligned}
    \frac{1}{N}\sum_{i,j=1}^{N}   f(\x_i,\x_j) v_{ic}v_{jc}= N\int \dd \x \dd\y~\rho_c( \x)\rho_c(\y)f(\x,\y)\ ,
\end{aligned}
\end{equation}
as can be easily verified via direct substitution, and we enforce their definition in Eq. \eqref{eq:Z^n_starting}  via functional deltas as

\begin{equation}
\begin{aligned}
    1=\int \{\mathcal{D}\rho_c\}\{\mathcal{D}\hat \rho_c\} \exp\left\{\ii N\sum_{c=1}^n\int \dd\x~\rho_c(\x) \hat \rho_c(\x)-\ii \sum_{c=1}^n\sum_{i=1}^N \hat \rho_c(\x_i)v_{ic}\right\} \ ,
\end{aligned}
\end{equation}
with the notation $\{\mathcal{D}\rho_c\}=\prod_c \mathcal{D}\rho_c$, $\{\mathcal{D}\hrho_c\}=\prod_c \mathcal{D}\hrho_c$. We can therefore rewrite the averaged replicated partition function (ignoring irrelevant constants whose logarithm vanishes in Eq.~\eqref{eq:lambda_max_limits}) as

\begin{equation}
\begin{aligned}\label{eq:Z^n_for_T(u)}
    \langle [Z_N(\beta)]^n \rangle \propto \int \dd\boldsymbol{\lambda}&\{\mathcal{D}\rho_c\}\{\mathcal{D}\hat \rho_c\} \exp\left\{\ii \frac{\beta}{2}N\sum_{c=1}^n\lambda_c+\frac{\beta}{2}N\sum_{c=1}^n\int \dd\x \dd\y~\rho_c(\x)\rho_c(\y)f(\x,\y)+\ii N\sum_{c=1}^n\int \dd\x~\rho_c(\x) \hat \rho_c(\x)\right\} \times \\
    & \times\int \prod_{c=1}^{n} \dd\Vec{v}_c \prod_{i=1}^N\dd\x_i~P(\x_1,\dots,\x_N) \exp\left\{-\ii\frac{\beta}{2}\sum_{c=1}^n\sum_{i=1}^N \lambda_c v_{ic}^2-\ii\sum_{c=1}^n\sum_{i=1}^N \hat \rho_c(\x_i)v_{ic}\right\} \ ,
\end{aligned}
\end{equation}
where $\dd\bm \lambda=\prod_{c=1}^{n} \dd \lambda_c$. Although from now on we will assume that the $N$ vectors $\x_i$ are i.i.d, such that $P(\x_1,\dots,\x_N)=\prod_{i=1}^N P(\x_i)$, we will relax this assumption in Appendix~\ref{sec:appendix_B} and deal with one case with correlated disorder of Curie--Weiss type. For i.i.d. vectors, the last integral in Eq.~\eqref{eq:Z^n_for_T(u)} can be factorised into $N$ identical integrals 

\begin{equation}
\begin{aligned}
    I_N \coloneq  &\int \prod_{c=1}^{n} \dd\Vec{v}_c \prod_{i=1}^N\dd\x_i~P(\x_1,\dots,\x_N) \exp\left\{-\ii\frac{\beta}{2}\sum_{c=1}^n\sum_{i=1}^N \lambda_c v_{ic}^2-\ii\sum_{c=1}^n\sum_{i=1}^N \hat \rho_c(\x_i)v_{ic}\right\}\\
    &= \int \prod_{c=1}^{n} \dd\Vec{v}_c \int\prod_{i=1}^N\left[\dd\x_i~P(\x_i) \exp\left\{-\ii\frac{\beta}{2}\sum_{c=1}^n \lambda_c v_{ic}^2-\ii\sum_{c=1}^n \hat \rho_c(\x_i)v_{ic}\right\}\right]\\
    &=\left[\int \dd\x \dd\boldsymbol{v}~P(\x) \exp\left\{-\ii\frac{\beta}{2}\sum_{c=1}^n \lambda_c v_{c}^2-\ii\sum_{c=1}^n \hat \rho_c(\x)v_{c}\right\}\right]^N\\
    &=\exp\left\{N \log\left[\int \dd\x P(\x) \prod_{c=1}^n \int \dd v \exp\left\{-\ii \frac{\beta}{2}\lambda_c v^2 - \ii \hat\rho_c(\x)v\right\}\right] \right\}\ ,
\end{aligned}
\end{equation}
where $\dd\bm v=\prod_{c=1}^{n} \dd v_c$. This leads to the following expression for the averaged replicated partition function

\begin{equation}
\begin{aligned}\label{eq:Z_saddle_point}
    \langle [Z_N(\beta)]^n \rangle \propto \int &\dd\boldsymbol{\lambda} \{\mathcal{D}\rho_c\} \{\mathcal{D}\hat \rho_c\}~ \exp\left\{ N \mathcal{S}[\boldsymbol{\lambda},\{\rho_c\} ,\{\hat \rho_c\}]\right\}\ ,
\end{aligned}
\end{equation}
where the action is defined as $\mathcal{S}[\boldsymbol{\lambda},\{\rho_c\}, \{\hat \rho_c\}]=\mathcal{S}_1[\boldsymbol{\lambda}]+\mathcal{S}_2[\{\rho_c\}]+\mathcal{S}_3[\{\rho_c\},\{\hat \rho_c\}]+\mathcal{S}_4[\boldsymbol{\lambda},\{\hat \rho_c\}]$, with

\begin{equation}
\begin{aligned}\label{eq:S1_complete}
    \mathcal{S}_1[\boldsymbol{\lambda}]=\ii\frac{\beta}{2}\sum_{c=1}^n\lambda_c\ , 
\end{aligned}
\end{equation}

\begin{equation}
\begin{aligned}\label{eq:S2_complete}
    \mathcal{S}_2[\{\rho_c\}]=\frac{\beta}{2}\sum_{c=1}^n\int \dd\x \dd\y~\rho_c(\x)\rho_c(\y)f(\x,\y)\ ,
\end{aligned}
\end{equation}

\begin{equation}
\begin{aligned}\label{eq:S3_complete}
    \mathcal{S}_3[\{\rho_c\},\{\hat \rho_c\}]=\ii\sum_{c=1}^n\int \dd\x~\rho_c(\x) \hat \rho_c(\x)\ ,
\end{aligned}
\end{equation}

\begin{equation}
\begin{aligned}
    \mathcal{S}_4[\boldsymbol{\lambda},\{\hat \rho_c\}]&=
    \log\left[\int \dd\x P(\x) \prod_{c=1}^n \int \dd v \exp\left\{-\ii \frac{\beta}{2}\lambda_c v^2 - \ii \hat\rho_c(\x)v\right\} \right]\ .
\end{aligned}
\end{equation}
The form of $\langle [Z_N(\beta)]^n \rangle$ in Eq. \eqref{eq:Z_saddle_point} is amenable to a saddle-point evaluation for large $N$, which yields \footnote{As specified following Eq.~\eqref{eq:lambda_max_limits}, we are assuming that the limit $N\to\infty$ can be taken first, before the zero-temperature and replica limits.}

\begin{equation}
    \langle [Z_N(\beta)]^n \rangle \underset{N \to \infty}{\approx} \exp\left\{ N \mathcal{S}[\boldsymbol{\lambda^*},\{\rho^*_c\}, \{\hat \rho^*_c\}]\right\}\ ,
\end{equation}
where $[\boldsymbol{\lambda^*},\{\rho^*_c\}, \{\hat \rho^*_c\}]$ maximise the action $\mathcal{S}$. Before deriving the saddle point equations, we assume a replica-symmetric structure \cite{noemi,pedestrian}, such that any
dependence on the specific replica index $c \in \{1,\dots,n\}$ chosen disappears

\begin{equation}
\rho_c=\rho,\quad \hat\rho_c=\hat\rho,\quad \lambda_c=\lambda, \qquad \forall c=1,\dots,n\ .
\end{equation}
We also expand the functional $\mathcal{S}[\lambda,\rho,\hrho]$ to the first order in $n$ (all other terms will disappear in the $n \to 0$ limit, see Eq. \eqref{eq:lambda_max_limits}), which gives

\begin{equation}
\begin{aligned}\label{eq:S_1_on}
    \mathcal{S}_1[\lambda]=\ii\frac{\beta}{2}n\lambda\ ,
\end{aligned}
\end{equation}

\begin{equation}
\begin{aligned}
    \mathcal{S}_2[\rho]=\frac{\beta}{2}n\int \dd\x \dd\y~\rho(\x)\rho(\y)f(\x,\y)\ ,
\end{aligned}
\end{equation}

\begin{equation}
\begin{aligned}\label{eq:S_3_on}
    \mathcal{S}_3[\rho,\hat \rho]=\ii n\int \dd\x~\rho(\x) \hat \rho(\x)\ ,
\end{aligned}
\end{equation}

\begin{equation}
\begin{aligned}\label{eq:S_4_on}
    \mathcal{S}_4[\lambda,\hat \rho]&= \log\left[\int \dd\x P(\x) \left( \int \dd v \exp\left\{-\ii \frac{\beta}{2}\lambda v^2 - \ii \hat\rho(\x)v\right\} \right)^n \right]\\
    &=n\int \dd\x~P(\x) ~\left(\log \int \dd v \exp\left\{-\ii \frac{\beta}{2}\lambda v^2 - \ii \hat\rho(\x)v\right\} \right)+o(n)\ ,
\end{aligned}
\end{equation}

where we used that $\int \dd\x~P(\x)=1$ by normalisation. Neglecting terms in $S_4[\lambda,\hat \rho]$ that are irrelevant in the limit $n\to 0$, the saddle point equations then reduce to

\begin{equation}
\begin{aligned}\label{eq:saddle_point_rho}
    0=\left.\frac{\delta\mathcal{S}[\lambda,\rho,\hrho]}{\delta \rho(\x)}\right|_{\lambda^\star,\rho^\star,\hrho^\star} \iff 0=\beta  \int \dd\y~\rho^*(\y)f(\x,\y) + \mathrm{i} \hat \rho^*(\x)\ ,
\end{aligned}
\end{equation}

\begin{equation}
\begin{aligned}\label{eq:saddle_point_hrho}
    0=\left.\frac{\delta\mathcal{S}[\lambda,\rho,\hrho]}{\delta \hat \rho(\x)}\right|_{\lambda^\star,\rho^\star,\hrho^\star}\iff 0&=\ii  \rho^*(\x)-\ii P(\x)\frac{\int \dd v~ \exp\left\{-\ii \frac{\beta}{2}\lambda^* v^2 - \ii \hat\rho^*(\x)v\right\}v }{\int \dd v~ \exp\left\{-\ii \frac{\beta}{2}\lambda^* v^2 - \ii \hat\rho^*(\x)v\right\}}
    \\&=\ii \rho^*(\x)+\frac{\ii}{\beta \lambda^*}P(\x)\hat \rho^*(\x)\ ,
\end{aligned}
\end{equation}

\begin{equation}
\begin{aligned}\label{eq:saddle_point_lambda}
    0=\left.\frac{\partial\mathcal{S}[\lambda,\rho,\hrho]}{\partial\lambda}\right|_{\lambda^\star,\rho^\star,\hrho^\star}\iff 0&=\ii\frac{\beta}{2}-\ii\frac{\beta}{2} \int \dd\x~P(\x)\frac{\int \dd v~ \exp\left\{-\ii \frac{\beta}{2}\lambda^* v^2 - \ii \hat\rho^*(\x)v\right\}v^2 }{\int \dd v~ \exp\left\{-\ii \frac{\beta}{2}\lambda^* v^2 - \ii \hat\rho^*(\x)v\right\}}\\
    &=\ii\frac{\beta}{2}\left[ 1+\frac{\ii}{\beta \lambda^*}-\int \dd\x~P(\x)\frac{\hat \rho^*(\x)^2}{\beta^2 \lambda^{*2}}\right]\ ,
\end{aligned}
\end{equation}
where we have used again that $\int \dd\x~P(\x)=1$ and performed the elementary $v$-integrals. Note that the imaginary part of $\lambda$ must be negative, $\Im(\lambda)<0$, to ensure their convergence\footnote{In Eq.~\eqref{eq:lambda_introduction}, the auxiliary variable $\lambda$ is introduced as a real variable. In the saddle-point evaluation of the large-$N$ integral in Eq.~\eqref{eq:Z_saddle_point}, however, the contour must be analytically continued and deformed to an equivalent one passing through the relevant complex saddle. The condition $\Im(\lambda)<0$ selects the half-plane where the Gaussian integrals entering the saddle-point equations are convergent. This deformation assumes that no singularities or branch cuts of the analytically continued integrand, equivalently of the action $\mathcal S$ on the chosen branch, are crossed.}. The solution of Eqs. \eqref{eq:saddle_point_rho}-\eqref{eq:saddle_point_lambda}, that maximises the functional $\mathcal{S}[\lambda,\rho,\hat{\rho}]$, are therefore given by

\begin{equation}
\begin{aligned}\label{eq:start_hat}
    \hat \rho^*(\x)=\ii\beta \int \dd\y~\rho^*(\y)f(\x,\y)=\ii \beta \psi(\x)\ ,
\end{aligned}
\end{equation}

\begin{equation}
\begin{aligned}\label{eq:star_rho}
    \rho^*(\x)=-\frac{1}{\beta \lambda^*}P(\x)\hat \rho^*(\x)= \frac{P(\x)}{\ii \lambda^*}\psi(\x)\ ,
\end{aligned}
\end{equation}

\begin{equation}
\begin{aligned}\label{eq:star_lambda_completa}
    \lambda^{*2}+\frac{\ii}{\beta}\lambda^*+\int \dd\x~ P(\x)\psi(\x)^2 =0 \iff \lambda^*=\frac{\ii}{2\beta}\left[-1-\sqrt{1+4\beta^2 s}\right]\ ,
\end{aligned}
\end{equation}
where we have introduced

\begin{equation}
\begin{aligned}\label{eq:psi_def_from_rho}
    &\psi(\x) \coloneq \int \dd\y~\rho^*(\y)f(\x,\y)\ ,\\
    &s\coloneq\int \dd\x~P(\x) \psi(\x)^2\ ,
\end{aligned}
\end{equation}
with $\psi(\x)$ to be determined self-consistently. In Eq. \eqref{eq:star_lambda_completa} we have selected the negative root for $\lambda^*$ to ensure that $\Im(\lambda^*)<0$ when $\psi(\x)$ is real. The final step to compute $\langle \lambda_{\mathrm{max}} \rangle$ from Eq.~\eqref{eq:lambda_max_limits} is to insert Eqs.~\eqref{eq:start_hat}-\eqref{eq:star_lambda_completa} into Eqs.~\eqref{eq:S_1_on}-\eqref{eq:S_4_on}, divide by $\frac{\beta n}{2}$, and take the limit $\beta \to \infty$. To this end, we further expand the action $\mathcal{S}$, already expanded to first order in $n$, in powers of $\beta$. Taking into account that from Eq.~\eqref{eq:star_lambda_completa}

\begin{equation}
\begin{aligned}\label{eq:star_lambda_o(b)}
    \lambda^*=\frac{\ii}{2\beta}\left[-1-\sqrt{1+4\beta^2 s}\right]=-\ii\sqrt{s}+\mathcal{O}\left(\frac{1}{\beta}\right)\ ,
\end{aligned}
\end{equation}
we get after simple algebra

\begin{equation}
\begin{aligned}
    \frac{2\mathcal{S}_1[\lambda^*]}{\beta n}=\ii\lambda^*=\sqrt{s}+\mathcal{O}\left(\frac{1}{\beta}\right)\ ,
\end{aligned}
\end{equation}

\begin{equation}
\begin{aligned}\label{eq:S2_star_o(b)}
    \frac{2\mathcal{S}_2[\rho^*]}{\beta n}
    &=\int \dd\x \dd\y~\rho^*(\x)\rho^*(\y)f(\x,\y)=\sqrt{s}+\mathcal{O}\left(\frac{1}{\beta}\right)\ ,
\end{aligned}
\end{equation}

\begin{equation}
\begin{aligned}
    \frac{2\mathcal{S}_3[\rho^*,\hat \rho^*]}{\beta n}&=\frac{2\ii}{\beta}\int \dd\x~\rho^*(\x)\hat \rho^*(\x)=-2\sqrt{s}+\mathcal{O}\left(\frac{1}{\beta}\right)\ ,
\end{aligned}
\end{equation}

\begin{equation}
\begin{aligned}\label{eq:fine_curie_weiss}
    \frac{2\mathcal{S}_4[\hat \rho^*,\lambda^*]}{\beta n}&=\frac{2}{\beta}\int \dd\x~P(\x) ~\left(\log \int \dd v \exp\left\{-\ii \frac{\beta}{2}\lambda^* v^2 - \ii \hat\rho^*(\x)v\right\} \right)=\sqrt{s}+\mathcal{O}\left(\frac{1}{\beta}\right)\ .
\end{aligned}
\end{equation}
Therefore, via Eq. \eqref{eq:lambda_max_limits}, the average of the largest eigenvalue is given by

\begin{equation}
\begin{aligned}\label{eq:lambda_max_final_long}
    \langle \lambda_{\mathrm{max}}\rangle= \sqrt{s} = \ii \lambda^*\ .
\end{aligned}
\end{equation}
The average largest eigenvalue thus coincides with the Lagrange multiplier $\ii\lambda^*$ at the saddle point (as usually happens in this type of problems \cite{Aufiero_2stipendi, PP_susca}) and is fully specified by the function $\psi(\x)$ through Eq. \eqref{eq:psi_def_from_rho}. Using the definition of $\psi(\x)$ in Eq.~\eqref{eq:psi_def_from_rho} and directly integrating Eq.~\eqref{eq:star_rho}, we obtain the self-consistency equation
\begin{equation}
\begin{aligned}
    \sqrt{s}~\psi(\x)=\int \dd\y~P(\y)f(\x,\y) \psi(\y)\ ,
\end{aligned}
\end{equation}
where, using Eq.~\eqref{eq:star_lambda_o(b)}, we have directly substituted $\ii\lambda^* = \sqrt{s}$, neglecting terms of order $\mathcal{O}\left(\frac{1}{\beta}\right)$ that vanish in the limit $\beta \to \infty$. Introducing the integral operator $T_f$, defined by
\begin{equation}
\begin{aligned}\label{eq:operator_Tf_iid}
    (T_f \psi)(\x)\coloneq \int \dd\y~P(\y)f(\x,\y)\psi(\y)\ ,
\end{aligned}
\end{equation}
the self-consistency equation can then be written as
\begin{equation}\label{eq:Fredholm_problem}
\sqrt{s}\,\psi=T_f\psi \ .
\end{equation}
Therefore, the replica saddle-point equations are equivalent to a Fredholm eigenvalue problem for $T_f$. In particular, in the large-$N$ limit, the average largest eigenvalue of the matrix $X$ defined in Eq.~\eqref{eq:matrix_X_generic} coincides with the largest eigenvalue of $T_f$. For the Fredholm spectral problem to be well defined, some regularity conditions on the kernel are needed. First, we assume that
\begin{equation}
\int \dd\x\,\dd\y\,P(\x)P(\y)|f(\x,\y)|^2<\infty\ ,
\end{equation}
which ensures that $T_f$ is a Hilbert--Schmidt operator, and hence is bounded and compact on $L^2(P)$ \cite{ReedSimon1980}. Since $f$ is symmetric, $T_f$ is self-adjoint with respect to the scalar product
\begin{equation}
(\phi,\psi)_P \coloneqq \int \dd\x\,P(\x)\phi(\x)\psi(\x)\ ,
\end{equation}
and the spectral theorem for compact self-adjoint operators implies that the non-zero spectrum consists of real eigenvalues of finite multiplicity, with possible accumulation only at zero \cite{ReedSimon1980}.
These assumptions alone do not necessarily guarantee the existence of a
non-trivial eigenfunction at the upper edge of the spectrum, if the upper edge
coincides with zero \cite{ReedSimon1980}. A sufficient condition excluding this case is
\begin{equation}
f(\x,\y)\geq 0
\qquad P\otimes P\text{-almost everywhere},
\qquad
f(\x,\y)>0 \text{ on a set of positive } P\otimes P\text{ measure}\ ,
\end{equation}
where $P\otimes P$ denotes the product measure associated with two independent
draws from $P$, so that $P\otimes P$-almost everywhere means up to a set of
pairs $(\x,\y)$ of zero probability under independent sampling. Under this assumption, the upper edge of the spectrum is strictly positive and, since $T_f$ is compact and self-adjoint, it is an eigenvalue of $T_f$. The Fredholm equation therefore admits at least one non-trivial solution associated with the principal eigenvalue, which is precisely the limiting largest eigenvalue of the matrix problem, $\langle \lambda_{\max}\rangle=\sqrt{s}$. Moreover, in this case $T_f$ is positivity preserving, so that the associated
eigenfunction can be chosen to be non-negative $P$-almost everywhere, with its overall scale fixed by the definition of $s$ in Eq.~\eqref{eq:psi_def_from_rho}.
The previous condition guarantees the existence of a principal eigenfunction,
but does not in general imply that the principal eigenvalue is simple.
Simplicity follows under the stronger assumption that $T_f$ is positivity
improving. A sufficient condition for this property is
\begin{equation}\label{eq:positivity_improving}
f(\x,\y)>0
\qquad P\otimes P\text{-almost everywhere}\ .
\end{equation}
Under this additional condition\footnote{The condition in Eq.~\eqref{eq:positivity_improving} ensures that no positive-probability region of the sample space is left ``disconnected'' from the rest under the action of the kernel, which is what ultimately makes the leading eigenvalue simple. This is the continuum analogue of the Perron--Frobenius irreducibility condition for the finite-$N$ matrix in Eq.~\eqref{eq:matrix_X_generic}, namely that the non-negative matrix cannot be decomposed into independent blocks.\label{fn:ftnote_2}}, Jentzsch's theorem, or more generally the Krein--Rutman theorem, implies that the principal eigenvalue is simple, while the corresponding eigenfunction is strictly positive $P$-almost everywhere and unique up to multiplication by a positive constant \cite{Jentzsch1912,Schaefer1974,Zaanen1997,KreinRutman1962}. Since $T_f$ is self-adjoint, eigenfunctions associated with distinct eigenvalues are orthogonal; hence no eigenfunction associated with a different eigenvalue can have a definite sign. Therefore, in this case, finding a ($P$-almost everywhere) strictly positive solution of the Fredholm equation automatically identifies the principal eigenpair, and hence the limiting average largest eigenvalue $\langle\lambda_{\max}\rangle=\sqrt{s}$.
Beyond this general formulation, it is not possible to make further analytical progress without specifying the kernel $f$, together with the distribution $P$ when needed. 

In the remainder of the main text, we specialise to the quadratic distance kernel 
\begin{equation}
\begin{aligned}\label{eq:quadratic_distance_kernel}
    f(\x,\y)=||\x-\y||^2=||\x||^2+||\y||^2-2\x \cdot \y\ ,
\end{aligned}
\end{equation}
where $\x \cdot \y$ denotes the standard dot product for $d$-dimensional vectors. For this kernel, the above sufficient conditions are trivially verified, provided the fourth moment of $P$ is finite. Moreover, the corresponding Fredholm equation admits a finite-dimensional reduction, which allows us to determine both the largest eigenvalue and the associated top-eigenvector statistics explicitly, since the eigenfunction $\psi(\x)$ defined in Eq.~\eqref{eq:psi_def_from_rho} takes the form
\begin{equation}
\begin{aligned}\label{eq:psi_from_ABC}
    \psi(\x)
    &= \int \dd\y~\rho^*(\y)\|\x-\y\|^2 = A\|\x\|^2+\B\cdot\x+C\ ,
\end{aligned}
\end{equation}
where the $d+2$ coefficients
\begin{equation}
\begin{aligned}\label{eq:ABC_def_from_rho}
    &A \coloneq \int \dd\y~\rho^*(\y)\ ,\\
    &B_{\ell} \coloneq -2 \int \dd\y~\rho^*(\y)~y_\ell\ ,\quad \quad \forall \ell=1,\dots,d\ ,\\
    &C \coloneq \int \dd\y~\rho^*(\y)~\|\y\|^2\ ,\\
\end{aligned}
\end{equation}
are fixed self-consistently by inserting Eq.~\eqref{eq:psi_from_ABC} into the general Fredholm equation, Eq.~\eqref{eq:Fredholm_problem}. This yields\footnote{In Eq. \eqref{eq:ABC_sistemone_generale} the averages are all with respect to the distribution $P(\x)$, i.e. $\langle g(\x)\rangle=\int \dd\x~P(\x)g(\x)$.}

\begin{equation}
\begin{cases}\label{eq:ABC_sistemone_generale}
\displaystyle
A = \frac{\sum_{\ell} B_\ell \langle x_\ell \rangle + C}
{\sqrt{s} - \sum_{\ell} \langle x_\ell^2 \rangle}\ , \\[1.2em]

\displaystyle
B_q = -2 \,
\frac{
A \sum_{\ell} \langle x_\ell^2 x_q \rangle
+ \sum_{\ell \ne q} B_\ell \langle x_\ell x_q \rangle
+ C \langle x_q \rangle
}{
\sqrt{s} + 2 \langle x_q^2 \rangle
}\ , \quad \forall q=1,\dots,d\ , \\[1.2em]

\displaystyle
C =
\frac{
A \sum_{\ell,q} \langle x_\ell^2 x_q^2 \rangle
+ \sum_{q} B_q \sum_{\ell} \langle x_\ell^2 x_q \rangle
}{
\sqrt{s} - \sum_{\ell} \langle x_\ell^2 \rangle
}\ .
\end{cases}
\end{equation}
Using Eqs. \eqref{eq:star_rho},\eqref{eq:psi_def_from_rho},\eqref{eq:star_lambda_o(b)},\eqref{eq:psi_from_ABC} and \eqref{eq:ABC_def_from_rho}  we also find the additional condition 

\begin{equation}
\begin{aligned}
s = \int \dd\x~P(\x) \psi(\x)^2 = \ii \lambda^* \int \dd\x~\rho^*(\x) \psi(\x) = 
\ii\,\lambda^\star \left[
2AC - \sum_{\ell=1}^d\frac{B_\ell^2}{2}
\right]=  \sqrt{s}\left[
2AC - \sum_{\ell=1}^d\frac{B_\ell^2}{2}
\right]+\mathcal{O}\left(\frac{1}{\beta}\right)\, 
\end{aligned}
\end{equation}
and therefore in the zero-temperature limit $\beta\to\infty$

\begin{equation}\label{eq:s=2ac-b2/2}
\sqrt{s} = \ii\,\lambda^\star = 2AC - \sum_{\ell}\frac{B_\ell^2}{2}  \, ,
\end{equation}
which also shows, via Eq.~\eqref{eq:lambda_max_final_long}, that 

\begin{equation}
\begin{aligned}\label{eq:lambda_final_short}
    \langle\lambda_{\mathrm{max}}\rangle=\sqrt{s}=2AC-\sum_{\ell=1}^d \frac{B^2_{\ell}}{2}\ .
\end{aligned}
\end{equation}
In the quadratic distance-kernel case, the average largest eigenvalue is therefore fully specified by the coefficients $(A,\B,C)$, obtained by solving the system of equations in Eq.~\eqref{eq:ABC_sistemone_generale}. This system depends on the distribution $P(\x)$ and on mixed correlators among the vector components up to fourth order. As follows from Eq.~\eqref{eq:s=2ac-b2/2}, the system to solve is nonlinear and generally admits multiple solutions, including the trivial one $(A,\B,C)=(0,\underline{0},0)$. In addition, some real solutions may yield $\sqrt{s}=2AC-\sum_{\ell=1}^d B_\ell^2/2<0$, which are unphysical since, by Eq.~\eqref{eq:s=2ac-b2/2}, this would imply $\Im (\lambda^*)>0$, and the integrals in Eqs.~\eqref{eq:saddle_point_hrho} and \eqref{eq:saddle_point_lambda} would diverge. Accordingly, we restrict to real $(A,\B,C)$ with $\sqrt{s}>0$. In Sections~\ref{sec:res_isotropic_dist} and \ref{sec:results_iid} we prove that such solution exists in any dimension $d$ for broad classes of distributions $P(\x)$, and determine the corresponding coefficients explicitly. For more general distributions $P(\x)$ outside these classes, closed-form solutions for the coefficients may be out of reach, but the system can always be solved numerically. In Section~\ref{sec:results_general} we indeed perform extensive numerical simulations, solving the system numerically and finding solutions that satisfy the above conditions in all cases considered.

As noted above, the system may admit multiple solutions. In particular, it is apparent from the structure of Eq.~\eqref{eq:ABC_sistemone_generale} that if $(A,\B,C)$ is a real solution, then $(-A,-\B,-C)$ is also a solution. Together with the condition $\sqrt{s}=2AC-\sum_{\ell=1}^d B_\ell^2/2>0$, which requires $A$ and $C$ to have the same sign, this implies that their overall sign is fixed up to a convention. We therefore select the solution with $A>0$ and $C>0$. As we will see in Section~\ref{sec:eigenvector}, this choice simply fixes the same (positive) sign for all components of the associated top eigenvector, in full agreement with the Perron–Frobenius theorem\footnote{Provided the points $\{\x_i\}_{i=1}^N$ are not identical, in the quadratic distance-kernel case the finite-$N$ matrix $X$ in Eq. \eqref{eq:matrix_X_generic} is nonnegative and irreducible, hence the Perron--Frobenius theorem applies. See also footnote \ref{fn:ftnote_2}.}. In Appendix~\ref{sec:appendix_A}, we provide an interpretation of the coefficients $(A,\B,C)$ in terms of an exact finite-rank decomposition of the spectral problem. Appendix~\ref{sec:appendix_C} illustrates the general Fredholm formulation on a Gaussian kernel, showing how the largest eigenvalue can be computed directly from the integral eigenvalue problem when no finite-dimensional reduction is available. For other choices of the kernel, closed-form solutions are not guaranteed and must be examined on a case-by-case basis. In the next section, we extend the analysis to the associated top eigenvector and derive the distribution of its components $T(u)$.

\newpage

\section{Density of the top eigenvector's components}\label{sec:eigenvector}

Let $\vec{v}^{\, (1)}$ denote the top eigenvector of the matrix $X$ defined in Eq. \eqref{eq:matrix_X_generic}, and let 

\begin{equation}
     T(u)=\big\langle \frac{1}{N}\sum_{i=1}^N\delta(u-v_i^{\, (1)})\big\rangle\ ,
\end{equation}
be the probability density of the top eigenvector's components. To find $T(u)$ we introduce the auxiliary quantity 

\begin{equation}\label{eq:T_tilde}
    \tilde T(u)=\big\langle \frac{1}{N}\sum_{i=1}^N\delta(u-v_i) \big\rangle_{P_{\beta,X}}\ ,
\end{equation}
where $\big\langle\  \cdot \ \rangle_{P_{\beta,X}}$ denotes the average with respect to the Gibbs-Boltzmann distribution of vectors $\vec v$ for a fixed instance of the matrix $X$

\begin{equation}
    P_{\beta,X}(\vec v)= \frac{\exp\left\{\frac{\beta}{2}(\vec v,X\vec v)\right\} \delta(||\vec v||^2-N)}{\int \dd\vec v^{\,'} \exp\left\{\frac{\beta}{2}(\vec{v}^{\,'}, X\vec v^{\,'})\right\} \delta(||\vec v^{\,'}||^2-N)} \ .
\end{equation}
Since the Gibbs measure will localise around $X$'s top eigenvector in the $\beta \to \infty$ limit, $\tilde T(u)$ gives the density of the top eigenvector component for a given $N \times N$ random matrix $X$. Defining an auxiliary partition function 

\begin{equation}
Z_{\varepsilon,\beta}(t, X; u)
\coloneq
\int \dd\vec{v}\;
\exp\left(
\frac{\beta}{2}\, (\vec{v},X\vec{v})
+ \beta t \sum_{i=1}^N \delta_\varepsilon(u - v_i)
\right)\,
\delta\left(\|\vec{v}\|^2 - N\right)\ ,
\end{equation}
where $\delta_{\varepsilon}$ is a smooth regularizer of the Dirac delta function, we can rewrite $\tilde T(u)$ in Eq. \eqref{eq:T_tilde} as

\begin{equation}
\tilde T(u)=
\lim_{\varepsilon \to 0^+}
\frac{1}{\beta N}
\frac{\partial}{\partial t}
\log  Z_{\varepsilon,\beta}(t, X; u) 
\Big|_{t=0}.
\end{equation}
Averaging over all possible realisations of the matrix $X$, sending $\beta \to \infty$ and using again the replica trick, we can write the density of the top eigenvector's components in the large-$N$ limit as

\begin{equation}\label{eq:Tu_starting_limits}
T(u)= 
\lim_{\beta \to \infty} \langle \tilde T(u) \rangle=
\lim_{N \to \infty}
\lim_{\beta \to \infty}
\lim_{\varepsilon \to 0^+}
\lim_{n \to 0}
\frac{1}{\beta N}
\frac{\partial}{\partial t}
\frac{1}{n}
\log \left\langle
\left[ Z_{\varepsilon,\beta}(t, X; u) \right]^n
\right\rangle
\Big|_{t=0}.
\end{equation}
As for Eq.~\eqref{eq:lambda_max_limits}, we assume that the large-$N$ limit can be performed first, so that the zero-temperature, regularisation, and replica limits are taken on the resulting saddle-point expression. This method has been employed to compute the density of the top eigenvector's components in several ensembles of sparse matrices \cite{Aufiero_2stipendi,PP_susca}. Doing the same computations of Section \ref{sec:eigenvalue}, we arrive to (see Eq. \ref{eq:Z^n_for_T(u)})

\begin{equation}
\begin{aligned}
    \left\langle
\left[ Z_{\varepsilon,\beta}(t, X; u) \right]^n
\right\rangle &\propto \int \dd\boldsymbol{\lambda}\{\mathcal{D}\rho_c\}\{\mathcal{D}\hat \rho_c\} \exp\left\{\ii\frac{\beta}{2}N\sum_{c=1}^n\lambda_c+\frac{\beta}{2}N\sum_{c=1}^n\int \dd\x \dd\y~\rho_c(\x)\rho_c(\y)f(\x,\y)+\ii N\sum_{c=1}^n\int \dd\x~\rho_c(\x) \hat \rho_c(\x)\right\} \times \\
    & \times\int \prod_{c=1}^{n} \dd\Vec{v}_c \prod_{i=1}^N \dd\x_i~P(\x_1,\dots,\x_N) \exp\left\{-\ii\frac{\beta}{2}\sum_{c=1}^n\sum_{i=1}^N \lambda_c v_{ic}^2-\ii\sum_{c=1}^n\sum_{i=1}^N \hat \rho_c(\x_i)v_{ic}+\beta t \sum_{c=1}^n\sum_{i=1}^N \delta_{\epsilon}(u-v_{ic})\right\} \ ,
\end{aligned}
\end{equation}
and assuming again that the vectors $\{\x_i\}_{i=1}^N$ are i.i.d., the integral in the last line factorises into $N$ identical integrals

\begin{equation}
\begin{aligned}
    \tilde I_N \coloneq  &\int \prod_{c=1}^{n} \dd\Vec{v}_c \prod_{i=1}^Nd\x_i~P(\x_1,\dots,\x_N) \exp\left\{-\ii\frac{\beta}{2}\sum_{c=1}^n\sum_{i=1}^N \lambda_c v_{ic}^2-\ii\sum_{c=1}^n\sum_{i=1}^N \hat \rho_c(\x_i)v_{ic}+\beta t \sum_{c=1}^n\sum_{i=1}^N \delta_{\epsilon}(u-v_{ic})\right\}\\
    &=\left[\int \dd\x \dd\boldsymbol{v}~P(\x) \exp\left\{-\ii\frac{\beta}{2}\sum_{c=1}^n \lambda_c v_{c}^2-\ii\sum_{c=1}^n \hat \rho_c(\x)v_{c}+\beta t \sum_{c=1}^n \delta_{\epsilon}(u-v_{c})\right\}\right]^N\\
    &=\exp\left\{N \log\left[\int \dd\x P(\x) \prod_{c=1}^n \int \dd v \exp\left\{-\ii \frac{\beta}{2}\lambda_c v^2 - \ii \hat\rho_c(\x)v+\beta t  \delta_{\epsilon}(u-v)\right\}\right] \right\}\ .
\end{aligned}
\end{equation}
Therefore we can write the average of the replicated $Z_{\varepsilon,\beta}(t, X; u)$ in a form amenable to a saddle point evaluation

\begin{equation}
\begin{aligned}\label{eq:Z_eigenvector_saddlepoint}
    \left\langle
\left[ Z_{\varepsilon,\beta}(t, X; u) \right]^n
\right\rangle\propto \int &\dd\boldsymbol{\lambda} \{\mathcal{D}\rho_c\} \{\mathcal{D}\hat \rho_c\}~ \exp\left\{ N \mathcal{S}[\boldsymbol{\lambda},\{\rho_c\} ,\{\hat \rho_c\};\varepsilon,t,u]\right\}\ ,
\end{aligned}
\end{equation}
where $\mathcal{S}[\boldsymbol{\lambda},\{\rho_c\} ,\{\hat \rho_c\};\varepsilon,t,u]=\mathcal{S}_1[\boldsymbol{\lambda}]+\mathcal{S}_2[\{\rho_c\}]+\mathcal{S}_3[\{\rho_c\},\{\hat \rho_c\}]+\mathcal{S}_4[\boldsymbol{\lambda},\{\hat \rho_c\},\varepsilon,t]$, with $\mathcal{S}_1,\mathcal{S}_2,\mathcal{S}_3$ given respectively by Eqs. \eqref{eq:S1_complete}-\eqref{eq:S3_complete} and 

\begin{equation}
\begin{aligned}
    S_4[\boldsymbol{\lambda},\{\hrho_c\},\epsilon,t]= \log\left[\int \dd\x P(\x) \prod_{c=1}^n \int \dd v \exp\left\{-\ii \frac{\beta}{2}\lambda_c v^2 - \ii \hat\rho_c(\x)v+\beta t  \delta_{\epsilon}(u-v)\right\}\right]\ .
\end{aligned}
\end{equation}
A saddle point evaluation of Eq. \eqref{eq:Z_eigenvector_saddlepoint} yields

\begin{equation}\label{eq:saddle_point_eigenvector}
    \left\langle
\left[ Z_{\varepsilon,\beta}(t, X; u) \right]^n
\right\rangle \underset{N \to \infty}{\approx} \exp\left\{ N \mathcal{S}[\boldsymbol{\lambda^*},\{\rho^*_c\}, \{\hat \rho^*_c\};\varepsilon,t,u]\right\}\ ,
\end{equation}
where $[\boldsymbol{\lambda^*},\{\rho^*_c\}, \{\hat \rho^*_c\}]$ maximise the action. Since the partial derivative with respect to $t$ in Eq. \eqref{eq:Tu_starting_limits} acts only on terms that explicitly depend on $t$, we can safely set $t=0$ in the saddle point equations. Therefore, assuming again a replica-symmetric structure and neglecting vanishing terms for $n \to 0$ in $\mathcal{S}$, the resulting saddle point equations reduce to those already solved in Section \ref{sec:eigenvalue}. Hence, $\lambda^*$, $\rho^*$ and $\hat\rho^*$ are still given by Eqs.~\eqref{eq:start_hat}-\eqref{eq:star_lambda_completa} and

\begin{equation}
\begin{aligned}\label{eq:S4_eigenvector_saddle_on}
    S_4[\hrho^*,\lambda^*,\epsilon,t]= n \int \dd\x~P(\x) \log \left(\int \dd v \exp\left\{-\ii \frac{\beta}{2}\lambda^* v^2 - \ii \hat\rho^*(\x)v+\beta t \delta_{\epsilon}(u-v)\right\}\right)+o(n)\ ,
\end{aligned}
\end{equation}
is the only term in Eq. \eqref{eq:saddle_point_eigenvector} that explicitly depends on $t$. Inserting Eqs. \eqref{eq:saddle_point_eigenvector} and \eqref{eq:S4_eigenvector_saddle_on} in Eq. \eqref{eq:Tu_starting_limits} yields

\begin{equation}
\begin{aligned}\label{eq:T(u)_suitable_for_iid}
T(u)&=\lim_{\beta \to \infty}\frac{1}{\beta}\lim_{n \to 0}\frac{\partial}{\partial t}\frac{1}{n}S_4[\hrho^*,\lambda^*,\epsilon,t]\Big|_{\epsilon=t=0}\\
&=\lim_{\beta \to \infty} \int \dd\x~P(\x) \frac{
\exp\!\left(
- \ii \frac{\beta}{2}\,\lambda^* u^2
- \ii\, \hrho^*(\x)\, u
\right)
}{
\int \dd v\,
\exp\!\left(
- \ii \frac{\beta}{2}\,\lambda^* v^2
- \ii\, \hrho^*(\x)\, v
\right)
}\\
&=\int \dd\x~P(\x) \delta\left(u-\frac{\psi(\x)}{\sqrt{s}}\right)\ ,
\end{aligned}
\end{equation}
where we have used Eqs.~\eqref{eq:start_hat} and \eqref{eq:star_lambda_o(b)}. Eq.~\eqref{eq:T(u)_suitable_for_iid} holds for a generic symmetric kernel function $f(\x,\y)$, where $\psi(\x)$ is the eigenfunction solving the Fredholm equation~\eqref{eq:Fredholm_problem} associated with the principal eigenvalue $\sqrt{s}$. This result shows that, in the large-$N$ limit, the components of the properly rescaled top eigenvector are distributed as $\psi(\x)/\sqrt{s}$, with $\x$ sampled according to $P(\x)$. To proceed explicitly, we now specialise again to the quadratic distance kernel in Eq.~\eqref{eq:quadratic_distance_kernel}. In this case, using Eq.~\eqref{eq:psi_from_ABC}, the general expression in Eq.~\eqref{eq:T(u)_suitable_for_iid} becomes

\begin{equation}
\begin{aligned}\label{eq:T(u)_suitable_for_iid_quadratic_kernel}
    T(u)=\int \dd\x~P(\x)\delta\left(u-\frac{A\|\x\|^2+\B\cdot\x+C}{\sqrt{s}}\right)\ ,
\end{aligned}
\end{equation}
where the coefficients $(A,\B,C)$ are obtained by solving the system in Eq.~\eqref{eq:ABC_sistemone_generale}, with $\sqrt{s}$ given by Eq.~\eqref{eq:s=2ac-b2/2}. Defining 

\begin{equation}
\begin{aligned}
h(\x)\coloneqq u-\frac{A||\x||^2+\B \cdot \x+C}{\sqrt{s}}
\end{aligned}
\end{equation}
we can rewrite $T(u)$ as a surface integral over the $(d-1)$ dimensional hypersurface defined by $h(\x)=0$

\begin{equation}\label{eq:T_over_h=0}
T(u)
=
\int \dd\x\, P(\x)\, \delta\!\big(h(\x)\big)
=
\int_{h(\x)=0}
\frac{P(\x)}{\|\nabla h(\x)\|}\, \dd\Sigma(\x)
=
\sqrt{s}\,
\int_{h(\x)=0}
\frac{P(\x)}{\|2A\x + \B\|}\, \dd\Sigma(\x)\ ,
\end{equation}
where $\dd \Sigma$ is the corresponding surface measure. Following the discussion at the end of Section \ref{sec:eigenvalue}, we can assume that $A$ and $\sqrt s$ are strictly positive real numbers, therefore the level set corresponding to $h(\x)=0$ 

 \begin{equation}
h(\x)=0 \iff \left\|\x+\frac{\B}{2A}\right\|^2=\frac{u\sqrt{s}-C}{A}+\frac{||\B||^2}{4A^2}\ ,
\end{equation}
is a $(d-1)$ sphere centered in $\x_c \coloneq -\frac{\B}{2A}$ and with radius $R(u)$ that can be written, using Eq. \eqref{eq:s=2ac-b2/2}, as

\begin{equation}\label{eq:R(u)_general}
R^2(u)=\frac{u\sqrt{s}-C}{A}+\frac{||\B||^2}{4A^2}=\frac{\sqrt{s}(2Au-1)}{2A^2}\ . 
\end{equation}
Therefore, we can write the surface integral in Eq. \eqref{eq:T_over_h=0} as an integral over the unit $(d-1)$ sphere $S_{d-1}=\{\vec{n}\in\R^d:||\vec n||=1\}$

\begin{equation}\label{eq:T_over_Sd-1=0}
\begin{aligned}
T(u)&=
\sqrt{s}\,
\int_{h(\x)=0}
\frac{P(\x)}{\|2A\x + \B\|}\, \dd\Sigma(\x)=\sqrt{s}\, \int_{||\x-\x_c||=R(u)}
\frac{P(\x)}{\|2A\x + \B\|}\, \dd\Sigma(\x)\\
&=\frac{\sqrt{s}}{2A}R(u)^{d-2}\left[\int_{S_{d-1}}P\left(-\frac{\B}{2A}+R(u)\vec n\right) \dd\Omega_{d-1}\right]\Theta(2Au-1)\ ,
\end{aligned}
\end{equation}
where we have used that

\begin{equation}
2A\x+\B=2A\left(\x+\frac{\B}{2A}\right)=2A(\x-\x_c)\iff ||2A\x+\B||=2A||\x-\x_c|| \ , 
\end{equation}
and that on a  $(d-1)$ sphere of radius $R(u)$ we have $\dd\Sigma(\x)=R(u)^{d-1}\dd\Omega_{d-1}$, where $\dd\Omega_{d-1}$ is the solid angle element. As a final remark we have that $T(u) \propto \Theta(R^2(u))$, because if $R(u)$ is a complex number then the equation $h(\x)=0$ has no solution in $\R^d$, and therefore $T(u)=0$ from Eq. \eqref{eq:T_over_h=0}. Recalling that $\sqrt{s}$ and $A$ are strictly positive, if follows from Eq. \eqref{eq:R(u)_general} that $\Theta(R^2(u))=\Theta(2Au-1)=\Theta(u-\frac{1}{2A})$, confirming that all components of the top eigenvector are positive. In the next section, we present the results for the top eigenpair, comparing analytical predictions with numerical simulations for different classes of distributions, including isotropic, i.i.d.\ symmetric, and more general cases.

\section{Results and numerical simulations for the quadratic distance kernel}\label{sec:results}
In this section we present the results for the top eigenpair in the quadratic distance kernel case and compare them with numerical simulations. To test the predictions obtained in Sections~\ref{sec:eigenvalue} and \ref{sec:eigenvector}, we perform extensive numerical simulations by directly diagonalising matrices of the form given in Eq.~\eqref{eq:matrix_X_generic}, with $f(\x_i,\x_j)=\|\x_i-\x_j\|^2$ and vectors $\{\x_i\}_{i=1}^N$ drawn i.i.d.\ from a distribution $P(\x)$. The theoretical predictions require solving the nonlinear system in Eq.~\eqref{eq:ABC_sistemone_generale} under the constraint $\sqrt{s}=2AC-\sum_{\ell=1}^d B_\ell^2/2>0$. In Section~\ref{sec:eigenvalue} we anticipated that, for certain distributions $P(\x)$, this constrained system can be solved analytically, yielding explicit expressions for the coefficients $(A,\B,C)$. In the following sections we present two such cases, namely isotropic distributions and i.i.d.\ symmetric distributions, for which the coefficients can be determined in closed form. We then turn to more general distributions, where the system is solved numerically. To perform the simulations and compare them with the replica method predictions, we fix a number of parameters common to all plots and simulations. In particular, we consider matrices of size $N=1500$, and compute averages $\langle \lambda_{\max} \rangle$ and histograms for $T(u)$ over $100$ independent realisations of the matrix $X$. Following the discussion at the end of Sections~\ref{sec:eigenvalue} and \ref{sec:eigenvector}, we select the solution of Eq.~\eqref{eq:ABC_sistemone_generale} with $A>0$, so that $T(u)\propto \Theta(2Au-1)=\Theta(u-1/(2A))$. Accordingly, for each realisation we fix the sign of the top eigenvector so that its components are consistently taken to be positive before constructing the histograms, as ensured by the Perron–Frobenius theorem.

\subsection{Isotropic distributions}\label{sec:res_isotropic_dist}
If we assume that the distribution of each vector is invariant under rotations, i.e. $P(\x)=P(||\x||)$, the system of equations in Eq. \eqref{eq:ABC_sistemone_generale} can be easily solved. In fact isotropy directly implies that:

\begin{enumerate}
    \item $\langle x_\ell \rangle = 0 \quad \forall \ell=1,\dots,d$
    \item $\langle x_\ell x_q \rangle = \sigma^2 \delta_{\ell q} \quad \forall \ell=1,\dots,d$
    \item  $\langle x_\ell x_q x_m x_n\rangle=M(\delta_{\ell q}\delta_{mn}+\delta_{\ell m}\delta_{qn}+\delta_{\ell n}\delta_{qm})$
    \item If $\sum_{\ell=1}^d \alpha_\ell$ is odd then $\langle x_1^{\alpha_1}\cdots x_d^{\alpha_d} \rangle=0$
\end{enumerate}

where $M$ and $\sigma$ are positive real constants that depend on the underlying distribution $P(\x)$. Using the third property above, all fourth-order moments can be computed directly, in particular 

\begin{itemize}
    \item $\langle x_\ell^4 \rangle = 3M \quad \forall \ell=1,\dots,d$
    \item $\langle x_\ell^2 x_q^2 \rangle = M \quad \forall \ell,q=1,\dots,d$ with $\ell \neq q$
    \item $\langle||\x||^4\rangle=\sum_{\ell,m=1}^d \langle x_\ell^2 x_m^2 \rangle=d(d+2)M\ .$
\end{itemize}

The system in Eq. \eqref{eq:ABC_sistemone_generale} therefore reduces to

\begin{equation}
\begin{cases}\label{eq:ABC_sistemone_isotropia}
\displaystyle
A = \frac{C}
{\sqrt{s} - d \sigma^2}\ , \\[1.2em]

\displaystyle
B_q = 0\ , \qquad \qquad \forall q=1,\dots,d \\[1.2em]

\displaystyle
C =
\frac{
A d(d+2)M
}{
\sqrt{s} - d \sigma^2
}\ ,
\end{cases}
\end{equation}

which directly implies that

\begin{equation}
    \begin{aligned}
        (\sqrt s - d \sigma^2)^2=d(d+2)M\iff \sqrt s= d \sigma^2+\sqrt{d(d+2)M}\ ,
    \end{aligned}
\end{equation}

where we have selected only the positive root to ensure that $\sqrt s >0$. Selecting the solution with $A,C>0$ we get

\begin{equation}
\begin{cases}
\displaystyle
A =  \frac{1}{\sqrt 2} \sqrt{1+ \sqrt{\frac{d}{d+2}} \frac{\sigma^2}{\sqrt{M}} }\ , \\[1.2em]

\displaystyle
C =
A \sqrt{d(d+2)M}
=
 \sqrt{\frac{d(d+2)M}{2}} \sqrt{1+ \sqrt{\frac{d}{d+2}} \frac{\sigma^2}{\sqrt{M}}}
\ .
\end{cases}
\end{equation}

The average largest eigenvalue can then be computed from Eq.~\eqref{eq:lambda_final_short}, yielding

\begin{equation}
    \begin{aligned}\label{eq:lambda_max_isotropic}
        \langle \lambda_{max} \rangle = \sqrt s= 2AC= d \sigma^2+\sqrt{d(d+2)M}\ .
    \end{aligned}
\end{equation}

For the density of the top eigenvector's components, the angular integral in Eq. \eqref{eq:T_over_Sd-1=0} is now trivial. In fact considering that in the isotropic case $\B=0$ by Eq. \eqref{eq:ABC_sistemone_isotropia}, and that $P(\x)$ is constant over the sphere we get

\begin{equation}
\begin{aligned}\label{eq:Tu_isotropic}
T(u)&= \frac{\pi^{\frac{d}{2}} \sqrt{s}}{A\Gamma(\frac{d}{2})}R(u)^{d-2}P(R(u))\Theta\left(u-\frac{1}{2A}\right)\ ,
\end{aligned}
\end{equation}

where $R(u)$ is given by Eq. \eqref{eq:R(u)_general}.
To test the analytical results obtained via the replica method, we consider three isotropic distributions: the uniform distribution on the sphere, the uniform distribution in the ball, and the generalized radial exponential. Their explicit form, together with the corresponding parameters $\sigma^2$ and $M$, are reported in Table~\ref{tab:isotropic}. 

\begin{figure}[htbp]
    \centering
    \includegraphics[width=0.85\textwidth]{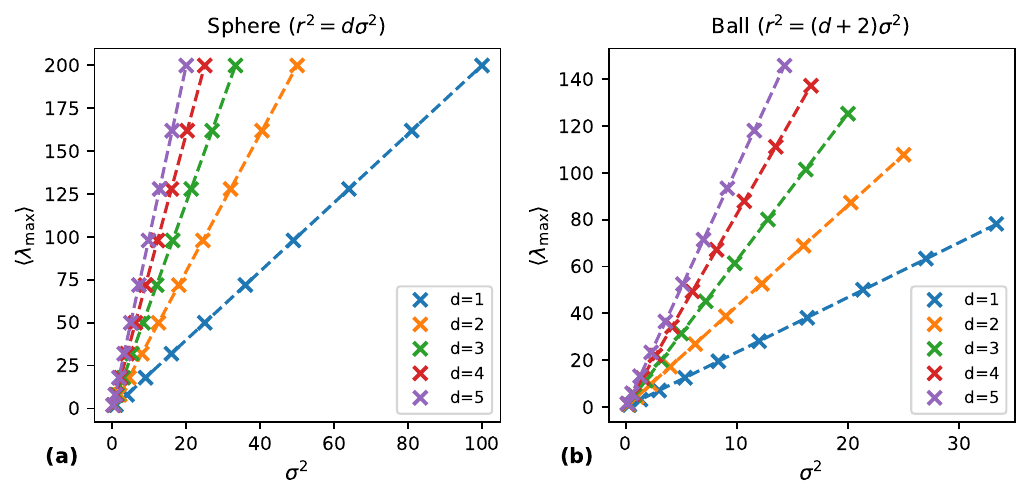}
    \caption{Scaling of $\langle\lambda_{\max}\rangle$ for uniform isotropic distributions. Comparison between the theoretical prediction given by Eq.~\eqref{eq:lambda_max_isotropic} (dashed lines) and the values obtained via direct numerical diagonalisation (colored crosses), as a function of $\sigma^2$ for different dimensions $d$ (colors). Panels \textbf{(a)} and \textbf{(b)} correspond to the uniform distribution on the sphere and in the ball, respectively. For every dimension $d$, the values of $\sigma^2$ used in the plots are obtained by varying the radius $r$ of the sphere and the ball from $1$ to $10$. Numerical results are obtained for matrices of size $N=1500$ and averaged over $100$ independent realisations.}
    \label{fig:lambda_sphere_ball}
\end{figure}

\begin{figure}[htbp]
    \centering
    \includegraphics[width=\textwidth]{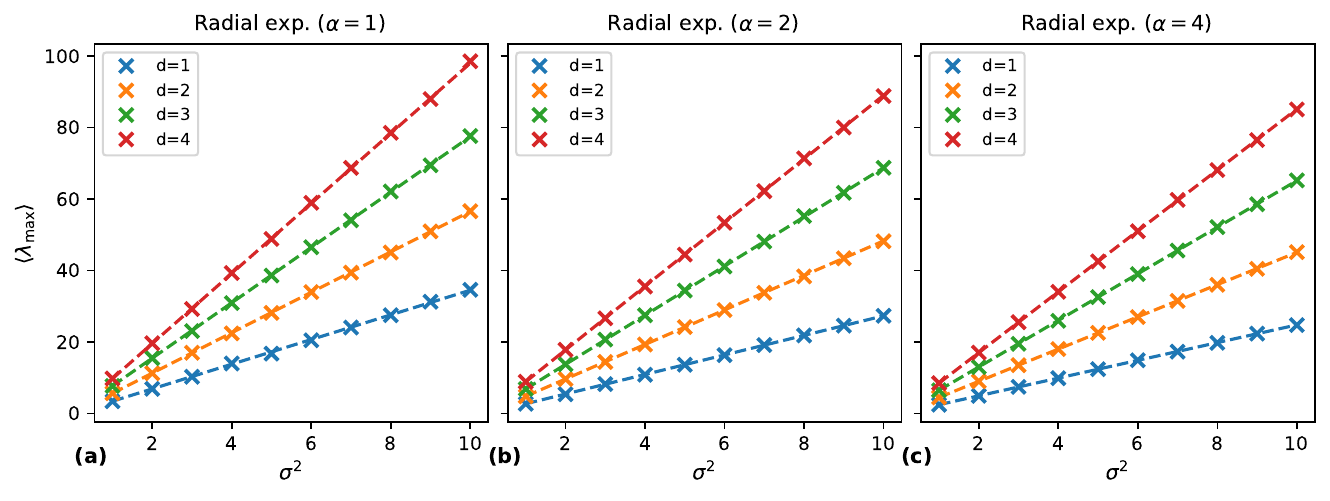}
    \caption{Scaling of $\langle\lambda_{\max}\rangle$ for the isotropic generalized exponential distributions. Comparison between the theoretical prediction given by Eq.~\eqref{eq:lambda_max_isotropic} (dashed lines) and the values obtained via direct numerical diagonalisation (colored crosses), as a function of $\sigma^2$ for different dimensions $d$ (colors). Panels \textbf{(a)}, \textbf{(b)}, and \textbf{(c)} correspond to different values of the shape  parameter $\alpha=1,2,4$, with $\alpha=1$ corresponding to the isotropic Laplace distribution and $\alpha=2$ recovering the centered Gaussian distribution. Numerical results are obtained for matrices of size $N=1500$ and averaged over $100$ independent realisations.}    
    \label{fig:lambda_exponential}
\end{figure}

In Fig.~\ref{fig:lambda_sphere_ball} and Fig.~\ref{fig:lambda_exponential} we compare the theoretical prediction for $\langle \lambda_{\max} \rangle$, given by Eq.~\eqref{eq:lambda_max_isotropic} (dashed lines), with the values obtained via direct diagonalisation (colored crosses), as a function of $\sigma^2$ and for different dimensions $d$ depicted with different colors. In Fig.~\ref{fig:lambda_sphere_ball}, panels \textbf{(a)} and \textbf{(b)} correspond to the uniform distribution on the sphere and in the ball, respectively, where the values of $\sigma^2$ used in the plots are obtained by varying the radius $r$ of the sphere and of the ball from $1$ to $10$. In Fig.~\ref{fig:lambda_exponential}, panels \textbf{(a)}, \textbf{(b)}, and \textbf{(c)} correspond to different values of the shape parameter $\alpha=1,2,4$, with $\alpha=1$ corresponding to the isotropic Laplace distribution and $\alpha=2$ recovering the centered Gaussian distribution. In all cases we observe perfect agreement between theory and simulations. In particular, $\langle \lambda_{\max} \rangle$ is linear in $\sigma^2$, in agreement with Eq.~\eqref{eq:lambda_max_isotropic}, since in general $M \propto \sigma^4$. 

\begin{table}[t]
\centering
\renewcommand{\arraystretch}{1.9}
\setlength{\tabcolsep}{14pt}

\begin{tabular}{|l|c|c|c|}
\hline
\rule{0pt}{3.2ex}
& $P(\x)$
& $\sigma^2$
& $M$ \\
\hline

\rule{0pt}{4.8ex}
Uniform on the sphere 
& $\displaystyle \frac{\Gamma(d/2)}{2\pi^{d/2}\,r^{d-1}} \delta(r-||\x||)$
& $\displaystyle \frac{r^2}{d}$
& $\displaystyle \frac{r^4}{d(d+2)}$
\\[1.2ex]
\hline

\rule{0pt}{4.8ex}
Uniform in the ball 
& $\displaystyle \frac{\Gamma(d/2)d}{2\pi^{d/2}\,r^d}\,\Theta(r - ||\x||)$
& $\displaystyle \frac{r^2}{d+2}$
& $\displaystyle \frac{r^4}{(d+2)(d+4)}$
\\[1.2ex]
\hline

\rule{0pt}{4.8ex}
Radial exponential
& $\displaystyle 
\frac{\alpha\,b^{d/\alpha}\Gamma(d/2)}{2\pi^{d/2}\Gamma(d/\alpha)}
\,\exp\{-b||\x||^\alpha\}
$
& $\displaystyle 
\frac{b^{-2/\alpha}}{d}\,\frac{\Gamma((d+2)/\alpha)}{\Gamma(d/\alpha)}
$
& $\displaystyle 
\frac{b^{-4/\alpha}}{d(d+2)}\,\frac{\Gamma((d+4)/\alpha)}{\Gamma(d/\alpha)}
$
\\[1.2ex]
\hline
\end{tabular}
\caption{Isotropic distributions in dimension $d$: probability density $P(\x)$ and the corresponding parameters $\sigma^2$ and $M$. Here $r$ denotes the radius of the sphere or ball, while $b$ and $\alpha$ are the parameters of the generalized radial exponential. Setting $\alpha=2$ corresponds to the centered Gaussian distribution.}
\label{tab:isotropic}
\end{table}

\FloatBarrier

\begin{figure}[htbp]
    \centering
    \includegraphics[width=0.85\textwidth]{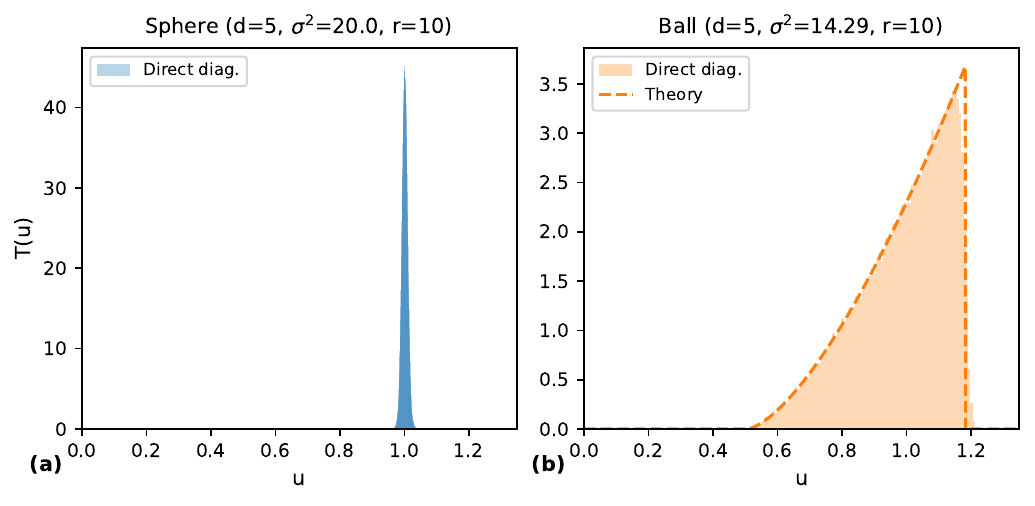}
    \caption{Density of the top eigenvector's components $T(u)$ for uniform isotropic distributions. Histograms obtained via direct numerical diagonalisation compared with the theoretical prediction given by Eq.~\eqref{eq:Tu_isotropic} (dashed lines). The dimension is fixed to $d=5$. Panel \textbf{(a)} corresponds to the uniform distribution on the sphere with radius $r=10$, for which the replica method predicts $T(u)=\delta(u-1)$, while panel \textbf{(b)} corresponds to the uniform distribution in the ball with the same radius. Numerical results are obtained for matrices of size $N=1500$ and averaged over $100$ independent realisations.}
    \label{fig:Tu_uniform_isotropic}
\end{figure}

\begin{figure}[htbp]
    \centering
    \includegraphics[width=\textwidth]{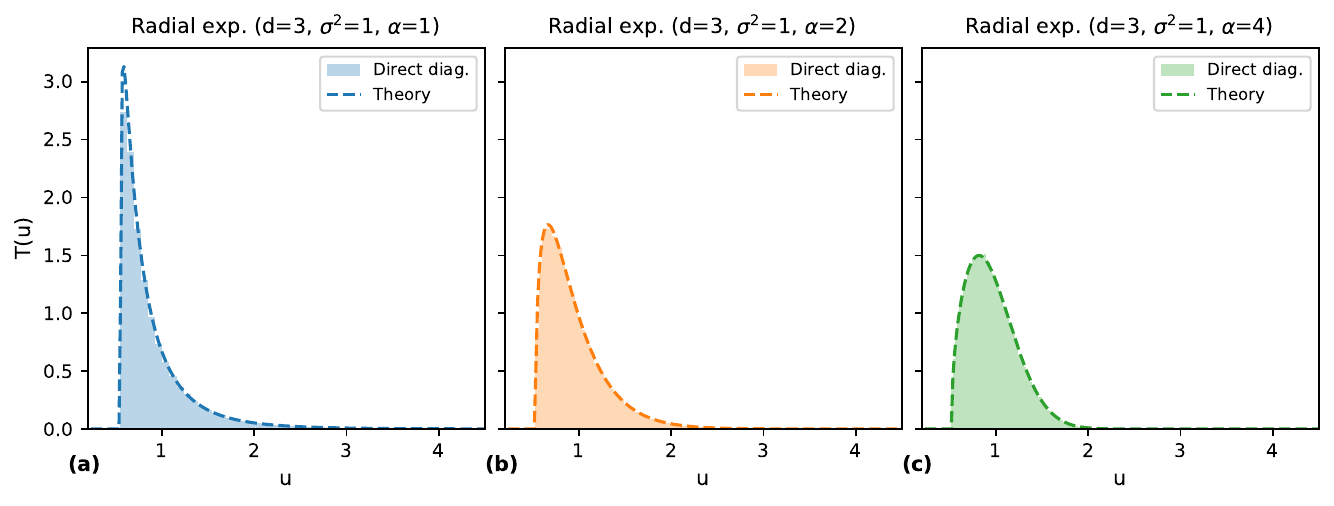}
    \caption{Density of the top eigenvector's components $T(u)$ for isotropic generalized exponential distributions. Histograms obtained via direct numerical diagonalisation compared with the theoretical prediction given by Eq.~\eqref{eq:Tu_isotropic} (dashed lines). The dimension is fixed to $d=3$ and $\sigma^2=1$. Panels \textbf{(a)}, \textbf{(b)}, and \textbf{(c)} correspond to different values of the shape parameter $\alpha=1,2,4$, with $\alpha=1$ corresponding to the isotropic Laplace distribution and $\alpha=2$ recovering the centered Gaussian distribution. Numerical results are obtained for matrices of size $N=1500$ and averaged over $100$ independent realisations.}  
    \label{fig:Tu_exponential_isotropic}
\end{figure}

We now consider the density of the top eigenvector's components $T(u)$. In Fig.~\ref{fig:Tu_uniform_isotropic} we fix the dimension to $d=5$ and plot the histogram of $T(u)$ obtained via direct numerical diagonalisation for the two uniform distributions. In panel \textbf{(a)}, corresponding to the uniform distribution on the sphere with radius $r=10$, Eq.~\eqref{eq:Tu_isotropic} predicts $T(u)=\delta(u-1)$, and we observe excellent agreement. In panel \textbf{(b)}, corresponding to the uniform distribution in the ball with radius $r=10$, the dashed line represents the replica prediction, again in very good agreement with the numerical results. In Fig.~\ref{fig:Tu_exponential_isotropic} we show the same comparison for the generalized radial exponential distribution, fixing the dimension to $d=3$ and $\sigma^2=1$. Histograms (direct diagonalisation) and dashed lines (replica prediction, Eq.~\eqref{eq:Tu_isotropic}) are shown for values of the shape parameter $\alpha=1,2,4$ in panels \textbf{(a)}, \textbf{(b)} and \textbf{(c)}, with excellent agreement in all cases.

\FloatBarrier

\subsection{Symmetric i.i.d. distribution}\label{sec:results_iid}

If we assume that the components of each vector are i.i.d, i.e. $P(\x)=\prod_{\ell = 1}^d P(x_\ell)$, in the system in Eq. \eqref{eq:ABC_sistemone_generale} all expectation values factorise, and it reduces to 

\begin{equation}
\begin{cases}\label{eq:ABC_sistemone_iid}
\displaystyle
A = \frac{\sum_{\ell} B_\ell \mu_1 + C}
{\sqrt{s} - d\mu_2}\ , \\[1.2em]

\displaystyle
B_q = -2 \,
\frac{
A [\mu_3+(d-1)\mu_2\mu_1]
+ \sum_{\ell \ne q} B_\ell \mu_1^2
+ C \mu_1
}{
\sqrt{s} + 2 \mu_2
}\ , \quad \forall q=1,\dots,d \\[1.2em]

\displaystyle
C =
\frac{
A [d(d-1)\mu_2^2+d\mu_4]
+ \sum_{q} B_q [(d-1)\mu_2\mu_1+\mu_3]
}{
\sqrt{s} - d\mu_2
}\ ,
\end{cases}
\end{equation}

where $\mu_n=\langle x^n \rangle$ is the $n$-th moment of the distribution. If we now additionally assume that the distribution is even, $P(x_\ell)=P(-x_\ell)$, all odd moments are zero and $\mu_2=\sigma^2$. Therefore the equations for the coefficients further simplify to

\begin{equation}
\begin{cases}\label{eq:ABC_sistemone_iid_simmetrico}
\displaystyle
A = \frac{C}
{\sqrt{s} - d\mu_2}\ , \\[1.2em]

\displaystyle
B_q = 0\ , \quad \forall q=1,\dots,d \\[1.2em]

\displaystyle
C =
\frac{
A [d(d-1)\mu_2^2+d\mu_4]
}{
\sqrt{s} - d\mu_2
}\ ,
\end{cases}
\end{equation}

which directly implies that

\begin{equation}
    \begin{aligned}
        (\sqrt s - d \mu_2)^2=d(d-1)\mu_2^2+d\mu_4\iff \sqrt s= d \mu_2+\sqrt{d(d-1)\mu_2^2+d\mu_4}\ ,
    \end{aligned}
\end{equation}

where again we have selected only the positive root to ensure that $\sqrt{s}>0$. Selecting the solution with $A,C>0$ we get

\begin{equation}
\begin{cases}
\displaystyle
A = \frac{1}{\sqrt 2} \sqrt{1+ \frac{d\mu_2}{\sqrt{d(d-1)\mu_2^2+d\mu_4}}} \ , \\[1.2em]

\displaystyle
C =
A \sqrt{d(d-1)\mu_2^2+d\mu_4}
=
\sqrt{\frac{d(d-1)\mu_2^2+d\mu_4}{2}} \sqrt{1+ \frac{d\mu_2}{\sqrt{d(d-1)\mu_2^2+d\mu_4}}}
\ ,
\end{cases}
\end{equation}

The average largest eigenvalue can then be computed from Eq.~\eqref{eq:lambda_final_short}, yielding

\begin{equation}
    \begin{aligned}\label{eq:lambda_max_idd_symm}
        \langle \lambda_{max} \rangle = \sqrt s=  d \mu_2+\sqrt{d(d-1)\mu_2^2+d\mu_4}=d \sigma^2+\sqrt{d(d-1)\sigma^4+d\mu_4}\ .
    \end{aligned}
\end{equation}

In general, Eqs.~\eqref{eq:ABC_sistemone_iid_simmetrico}-\eqref{eq:lambda_max_idd_symm} do not match the results obtained in Section~\ref{sec:res_isotropic_dist} for isotropic distributions, namely Eqs.~\eqref{eq:ABC_sistemone_isotropia}--\eqref{eq:lambda_max_isotropic}, except in the trivial case of dimension $d=1$, where the two descriptions coincide. More generally, isotropy does not imply independence of the variables, and the two sets of results agree only under the additional condition $\mu_4=3\mu_2^2 = 3\sigma^4$, which, in the isotropic case, corresponds to $M = \sigma^4$. This is precisely the multivariate symmetric Gaussian case, known to be the only isotropic distribution with independent components. 
For the density of the top eigenvector's components, the angular integral in Eq.~\eqref{eq:T_over_Sd-1=0} is not trivial. Even though Eq.~\eqref{eq:ABC_sistemone_iid_simmetrico} gives $\B=0$, as in the isotropic case, $P(\x)$ now is not constant over the sphere and does not factorise in spherical coordinates. Nonetheless, a more convenient approach is possible. Starting from Eq.~\eqref{eq:T(u)_suitable_for_iid_quadratic_kernel} with $\B=0$ and using Eq.~\eqref{eq:R(u)_general}, we obtain

\begin{equation}
\begin{aligned}\label{eq:T(u)_iid_colmodulo}
T(u)&=\int \dd\x~P(\x) \delta\left(u-\frac{A||\x||^2+C}{\sqrt{s}}\right)\\
&=\int \dd\x~ \prod_{\ell=1}^dP(x_\ell) \delta\left(\frac{A}{\sqrt{s}}(R^2(u)-||\x||^2)\right)\\
&=\frac{\sqrt{s}}{A}\int \dd\x~ \prod_{\ell=1}^dP(x_\ell) \delta(R^2(u)-||\x||^2)=\frac{\sqrt{s}}{A} f_{||\x||^2}(R^2(u))
\ .
\end{aligned}
\end{equation}

This shows that $T(u)$ is proportional to the probability density $f_{||\x||^2}$ of the random variable $||\x||^2$, evaluated at $R^2(u)$. Since the components $x_\ell$ are i.i.d., the random variable $||\x||^2=\sum_{\ell=1}^d x_\ell^2$ is a sum of i.i.d. random variables, and its distribution can be obtained by convolution. Denoting by $q$ the distribution of the squared components $x^2_\ell$, a change of variables in the probability distribution yields

\begin{equation}\label{eq:q_dist}
q(z)= \frac{P(\sqrt{z})+P(-\sqrt{z})}{2\sqrt{z}}\Theta(z)=\frac{P(\sqrt{z})}{\sqrt{z}}\Theta(z)\ ,
\end{equation}

where in the last equality we used that $P$ is even. It then follows that

\begin{equation}
f_{||\x||^2}(R^2(u)) = q^{*d}(R^2(u))=\int_0^\infty \dd z_1\dots\int_0^\infty \dd z_d ~ q(z_1)\dots q(z_d)\delta\left(R^2(u)-\sum_{\ell=1}^dz_\ell\right)\ ,
\end{equation}

where $q^{*d}$ denotes the $d$-fold convolution of $q$ with itself. Recalling Eq.~\eqref{eq:T(u)_iid_colmodulo}, we can write

\begin{equation}
\begin{aligned}\label{eq:T(u)_iid_final}
T(u)&=\frac{\sqrt{s}}{A} f_{||\x||^2}(R^2(u))\Theta(R^2(u))
=\frac{\sqrt{s}}{A} q^{*d}(R^2(u))\,\Theta(2Au-1)\ .
\end{aligned}
\end{equation}

To test the analytical predictions in the i.i.d. symmetric case, we consider a distribution for the components $x_\ell$ of the form

\begin{equation}\label{eq:p_gamma}
P(x)=\frac{|x|^{2\gamma-1}}{\Gamma(\gamma)\theta^{\gamma}}\exp\!\left(-\frac{x^2}{\theta}\right)\ ,
\end{equation}

with $\gamma>0$ and $\theta>0$, and first two non-zero moments

\begin{equation}
\mu_2=\sigma^2=\gamma\theta \ , \qquad \mu_4=\frac{\gamma+1}{\gamma}\sigma^4 \ .
\end{equation}

Using Eq.~\eqref{eq:q_dist}, it follows that the distribution of the squared components $x_\ell^2$ is a Gamma distribution with shape parameter $\gamma$ and scale parameter $\theta$

\begin{equation}
q(z)=\frac{z^{\gamma-1}}{\Gamma(\gamma)\theta^{\gamma}}\exp\!\left(-\frac{z}{\theta}\right)\Theta(z)\ .
\end{equation}

As a consequence, the random variable $||\x||^2=\sum_{\ell=1}^d x_\ell^2$ is Gamma distributed with shape parameter $d\gamma$ and scale parameter $\theta$, and its distribution evaluated at $R^2(u)$ is therefore

\begin{equation}
f_{||\x||^2}(R^2(u))=q^{*d}(R^2(u))=\frac{R(u)^{2d\gamma-2}}{\Gamma(d\gamma)\theta^{d\gamma}}\exp\!\left(-\frac{R^2(u)}{\theta}\right)\Theta(R^2(u))\ .
\end{equation}

\begin{figure}[htbp]
    \centering
    \includegraphics[width=\textwidth]{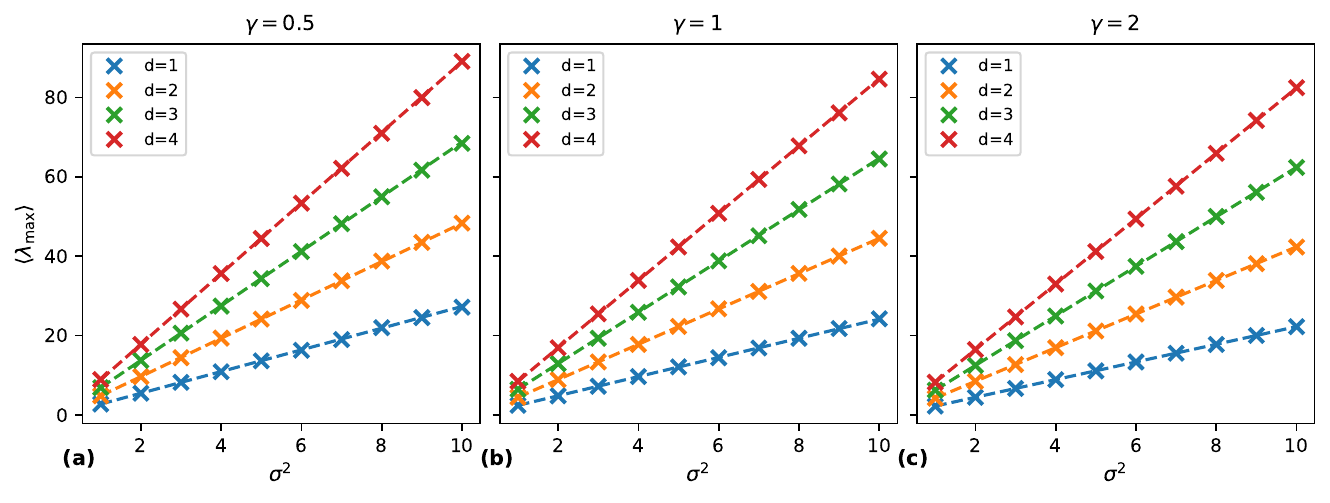}
    \caption{Scaling of $\langle \lambda_{\max} \rangle$ in the i.i.d.\ symmetric case for the distribution $P(x)$ defined in Eq.~\eqref{eq:p_gamma}. Comparison between the theoretical prediction given by Eq.~\eqref{eq:lambda_max_idd_symm} (dashed lines) and the values obtained via direct numerical diagonalisation (colored crosses), as a function of $\sigma^2$ for different dimensions $d$ (colors). Panels \textbf{(a)}, \textbf{(b)}, and \textbf{(c)} correspond to different values of the shape parameter $\gamma=0.5,1,2$, respectively, with $\gamma=0.5$ corresponding to the Gaussian case. Numerical results are obtained for matrices of size $N=1500$ and averaged over $100$ independent realisations.}
    \label{fig:lambda_iid_gamma}
\end{figure}

In Fig.~\ref{fig:lambda_iid_gamma} we compare the theoretical prediction for $\langle \lambda_{\max} \rangle$, given by Eq.~\eqref{eq:lambda_max_idd_symm} (dashed lines), with the values obtained via direct numerical diagonalisation (colored crosses), as a function of $\sigma^2$ and for different dimensions $d$ depicted with different colors. The three panels \textbf{(a)}, \textbf{(b)}, and \textbf{(c)} correspond to different values of the shape parameter $\gamma=0.5,1,2$, respectively, with $\gamma=0.5$ corresponding to the Gaussian case. In all cases we observe excellent agreement between theory and numerical results.

\begin{figure}[htbp]
    \centering
    \includegraphics[width=\textwidth]{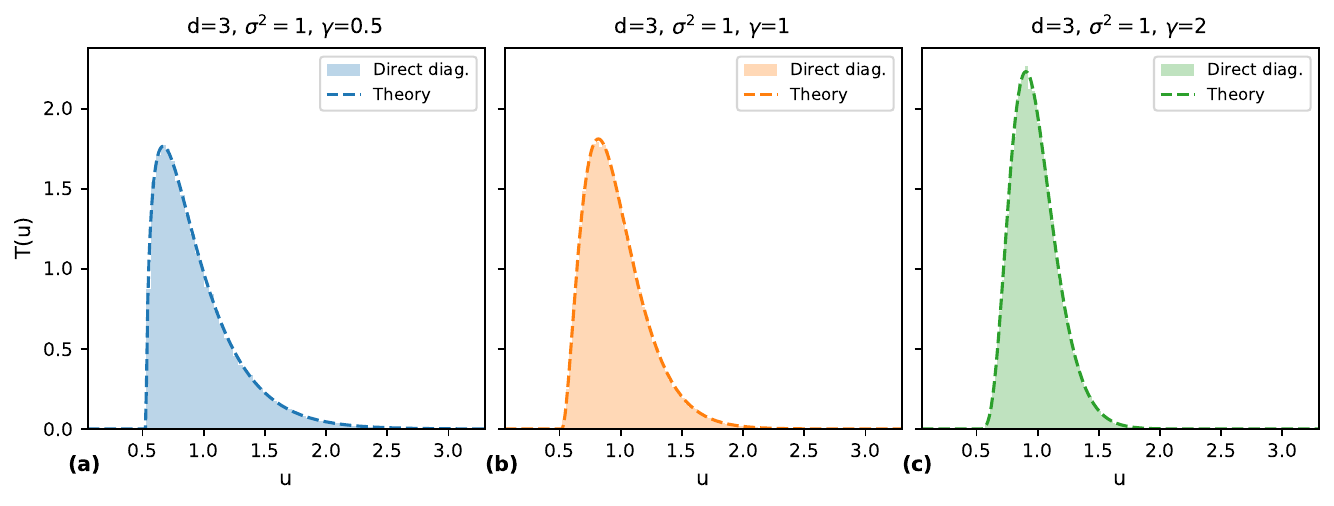}
    \caption{Density of the top eigenvector's components $T(u)$ in the i.i.d.\ symmetric case for the distribution $P(x)$ defined in Eq.~\eqref{eq:p_gamma}. Histograms obtained via direct numerical diagonalisation compared with the theoretical prediction given by Eq.~\eqref{eq:T(u)_iid_final} (dashed lines). The dimension is fixed to $d=3$ and $\sigma^2=1$. Panels \textbf{(a)}, \textbf{(b)}, and \textbf{(c)} correspond to different values of the shape parameter $\gamma=0.5,1,2$, respectively, with $\gamma=0.5$ corresponding to the Gaussian case. Numerical results are obtained for matrices of size $N=1500$ and averaged over $100$ independent realisations.}
    \label{fig:Tu_iid_gamma}
\end{figure}

We now consider the density of the top eigenvector's components $T(u)$. In Fig.~\ref{fig:Tu_iid_gamma} we fix the dimension to $d=3$ and $\sigma^2=1$, and compare the theoretical prediction given by Eq.~\eqref{eq:T(u)_iid_final} (dashed lines) with the results obtained via direct numerical diagonalisation (colored histograms). The three panels \textbf{(a)}, \textbf{(b)}, and \textbf{(c)} correspond to different values of the shape parameter $\gamma=0.5,1,2$, respectively. The agreement between theory and numerical results is again excellent across all cases.

\FloatBarrier

\subsection{Multivariate Gaussian distribution}\label{sec:results_general}
\begin{figure}[htbp]
    \centering
    \includegraphics[width=0.85\textwidth]{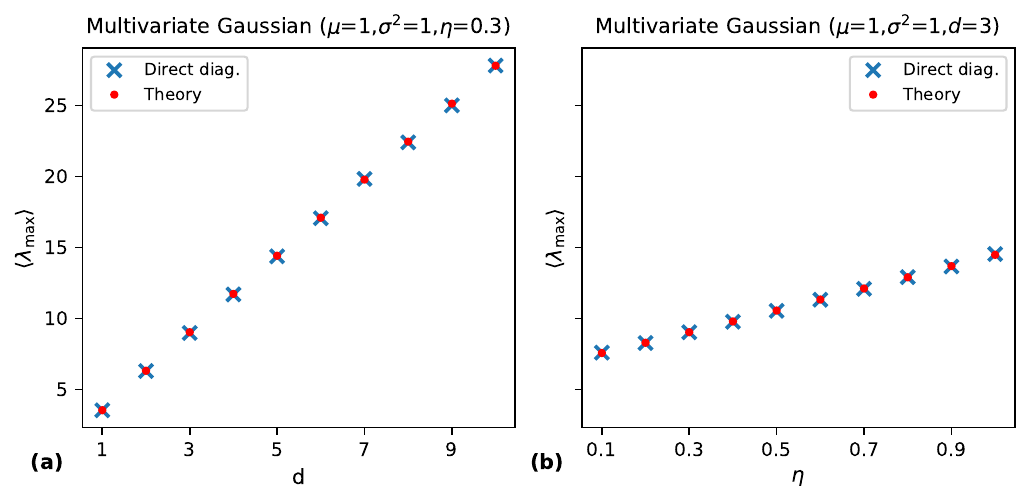}
    \caption{Scaling of $\langle \lambda_{\max} \rangle$ for a multivariate Gaussian distribution with $\underline{m}=\mu \underbar{1}$ and $\Sigma=\sigma^2 I_d + \eta \underbar{1}\underbar{1}^\top$. Comparison between the values obtained via direct numerical diagonalisation (blue crosses) and the replica predictions (red circles), computed by solving the system in Eq.~\eqref{eq:ABC_sistemone_generale} numerically. In panel \textbf{(a)} we fix $\mu=1$, $\sigma^2=1$, and $\eta=0.3$, and plot the results as a function of the dimension $d$. In panel \textbf{(b)} we fix $\mu=1$, $\sigma^2=1$, and $d=3$, and plot the results as a function of the correlation strength $\eta$. Numerical results are obtained for matrices of size $N=1500$ and averaged over $100$ independent realisations.}
    \label{fig:lambda_gaussian_general}
\end{figure}

We now consider a different setting that does not fall within the two special classes discussed above. In particular, we assume that each vector $\{\x_i\}_{i=1}^N$ is drawn independently from a $d$-dimensional multivariate Gaussian distribution with non-zero mean $\underline{m}$ and non-diagonal covariance matrix $\Sigma$. The moments up to fourth order entering the system in Eq.~\eqref{eq:ABC_sistemone_generale} are known explicitly and are given by the following set of equations

\begin{equation}
\begin{cases}
\langle x_\ell \rangle = m_\ell\ , \\[4pt]
\langle x_\ell x_q \rangle = m_\ell m_q + \Sigma_{\ell q}\ , \\[4pt]
\langle x_\ell^2 x_q \rangle = m_\ell^2 m_q + 2m_\ell \Sigma_{\ell q} + m_q \Sigma_{\ell\ell}\ , \\[4pt]
\langle x_\ell^2 x_q^2 \rangle = m_\ell^2 m_q^2 + m_\ell^2 \Sigma_{qq} + m_q^2 \Sigma_{\ell\ell} + 4m_\ell m_q \Sigma_{\ell q} + \Sigma_{\ell\ell}\Sigma_{qq} + 2\Sigma_{\ell q}^2\ .
\end{cases}
\qquad \forall \,\ell,q \in \{1,\dots,d\}
\end{equation}
In this case, obtaining a closed-form analytical solution for the coefficients $(A,\B,C)$ is in general intractable, and we therefore solve the system of equations in Eq.~\eqref{eq:ABC_sistemone_generale} numerically. Following the discussion in Section~\ref{sec:eigenvalue}, we impose the constraint $\sqrt{s}=2AC-\sum_{\ell=1}^d B_\ell^2/2>0$ and select the solution with $A,C>0$. The resulting values of $(A,\B,C)$ are then used to compute $\langle \lambda_{\max} \rangle$ from Eq.~\eqref{eq:lambda_final_short}, and this prediction is compared with the values obtained via direct numerical diagonalisation. As a concrete example, we take $\underline{m}=\mu \underbar{1}$ and $\Sigma=\sigma^2 I_d + \eta \underbar{1}\underbar{1}^\top$, where $I_d$ is the $d \times d$ identity matrix, $\mu,\sigma^2$ and $\eta$ are strictly positive constants, and $\underbar{1}$ denotes the $d$-dimensional vector with all components equal to one. In Fig.~\ref{fig:lambda_gaussian_general} we compare the values of $\langle \lambda_{\max} \rangle$ obtained via direct numerical diagonalisation (blue crosses) with the corresponding replica predictions (red circles). In panel \textbf{(a)} we fix $\mu=1$, $\sigma^2=1$, and $\eta=0.3$, and plot the results as a function of the dimension $d$. In panel \textbf{(b)} we fix $\mu=1$, $\sigma^2=1$, and $d=3$, and plot the results as a function of the correlation strength $\eta$. In both cases we observe very good agreement between direct numerical diagonalisation and the theoretical predictions. 

\FloatBarrier

\begin{figure}[htbp]
    \centering
    \includegraphics[width=0.8\textwidth]{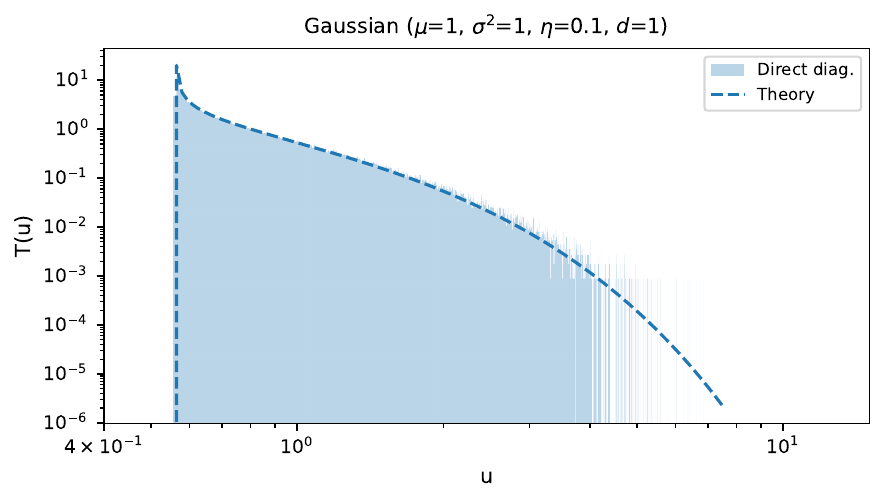}
    \caption{Density of the top eigenvector's components $T(u)$ for a Gaussian distribution with $\underline{m}=\mu \underbar{1}$ and $\Sigma=\sigma^2 I_d + \eta \underbar{1}\underbar{1}^\top$. Comparison between the results obtained via direct numerical diagonalisation (colored histogram) and the replica prediction given by Eq.~\eqref{eq:T_over_Sd-1=0} (dashed line). Parameters are fixed to $d=1$, $\mu=1$, $\sigma^2=1$, and $\eta=0.1$. The distribution is shown in double logarithmic scale. Numerical results are obtained for matrices of size $N=1500$ and averaged over $100$ independent realisations.}
    \label{fig:Tu_gaussian_general}
\end{figure}

In Fig.~\ref{fig:Tu_gaussian_general} we consider the density of the top eigenvector’s components $T(u)$. We fix $d=1$, $\mu=1$, $\sigma^2=1$, and $\eta=0.1$, and compare the results obtained via direct numerical diagonalisation (colored histogram) with the replica prediction given by Eq.~\eqref{eq:T_over_Sd-1=0} (dashed line). The distribution is shown in double logarithmic scale, and we observe again very good agreement between theory and numerical results.

\FloatBarrier

\subsection{Distributions with matching moments}
An important consequence of our results is that the coefficients $(A,\B,C)$ obtained by solving the system in Eq.~\eqref{eq:ABC_sistemone_generale} depend only on the moments of the distribution $P(\x)$ up to fourth order. Since the average largest eigenvalue $\langle \lambda_{\max} \rangle$ is entirely specified by these coefficients through Eq.~\eqref{eq:lambda_final_short}, it follows that distributions sharing the same first four moments yield identical theoretical predictions for $\langle \lambda_{\max} \rangle$, even if their higher-order statistics differ. To verify this prediction, we consider the one-dimensional case $d=1$ and compare two different even distributions sharing the same first four moments. In particular, we take a triangular distribution supported on $[-L,L]$ with mode at zero, and a mixture of two uniform distributions 

\begin{equation}\label{eq:P1-P2}
P_1(x)=
\begin{cases}
\dfrac{L-|x|}{L^2} & |x|\le L\\[6pt]
0 & |x|>L
\end{cases}
\qquad
P_2(x)=
\begin{cases}
\dfrac{1}{4a}+\dfrac{1}{4b} & |x|\le b\\[6pt]
\dfrac{1}{4a} & b<|x|\le a\\[6pt]
0 & |x|>a\ ,
\end{cases}
\end{equation}
with 
\begin{equation}
a^2=\frac{L^2}{2}\left(1+\frac{1}{\sqrt{3}}\right)\ , \qquad
b^2=\frac{L^2}{2}\left(1-\frac{1}{\sqrt{3}}\right)\ .
\end{equation}
It is straightforward to verify that these two distributions share the same moments up to fourth order,
\begin{equation}
\mu_1 = 0\ , \quad
\mu_2=\sigma^2 = \frac{L^2}{6}\ , \quad
\mu_3 = 0\ , \quad
\mu_4= \frac{L^4}{15}=\frac{36}{15}\sigma^4\ ,
\end{equation}

while higher-order moments differ. In this setting, since $d=1$ and the distributions are even, the results of Sections~\ref{sec:res_isotropic_dist} and \ref{sec:results_iid} coincide. From Eqs.~\eqref{eq:lambda_max_isotropic} and \eqref{eq:lambda_max_idd_symm}, we therefore obtain the same prediction for both distributions

\begin{equation}\label{eq:lambda_triangular=uniform}
\lambda_{\mathrm{\max}} = \sigma^2+\sqrt{\mu_4}=\left(1+\frac{6}{\sqrt{15}}\right)\sigma^2\ .
\end{equation}

\begin{figure}[htbp]
    \centering
    \includegraphics[width=0.6\textwidth]{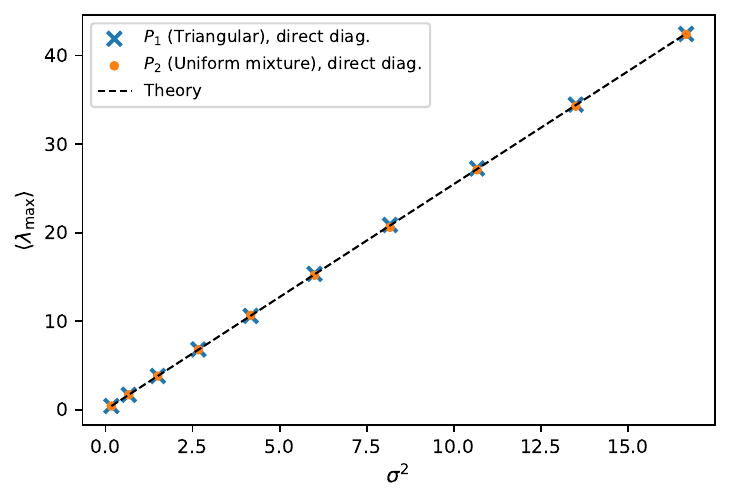}
    \caption{Scaling of $\langle \lambda_{\max} \rangle$ for two distributions sharing the same first four moments. Comparison between the theoretical prediction (black dashed line) and the values obtained via direct numerical diagonalisation (symbols), as a function of $\sigma^2$. Blue crosses correspond to the triangular distribution, while orange circles represent the uniform mixture defined in Eq.~\eqref{eq:P1-P2}. Numerical results are obtained for matrices of size $N=1500$ and averaged over $100$ independent realizations.}    
    \label{fig:lambda_triangular=uniform}
\end{figure}

In Fig.~\ref{fig:lambda_triangular=uniform} we compare the theoretical prediction in Eq.~\eqref{eq:lambda_triangular=uniform} (black dashed line) with the values of $\langle \lambda_{\max} \rangle$ obtained via direct numerical diagonalisation, as a function of $\sigma^2$, for the two distributions $P_1$ and $P_2$ in Eq. \eqref{eq:P1-P2}. The triangular distribution is represented by blue crosses, while the uniform mixture is shown with orange circles. In both cases, we observe very good agreement with the theoretical prediction. Moreover, the numerical results for the two distributions coincide within statistical fluctuations, confirming that $\langle \lambda_{\max} \rangle$ depends only on the first four moments of the distribution, as expected.

\FloatBarrier

\section{Conclusion}\label{sec:conclusion}
In this work, we investigated the spectral properties of symmetric Euclidean random matrices constructed from a set of $N$ random vectors in $\mathbb{R}^d$, drawn independently from a common distribution. In particular, we focused on the average largest eigenvalue $\langle \lambda_{\max} \rangle$ and on the density $T(u)$ of the components of the associated top eigenvector. Starting from a statistical mechanics formulation of the problem, we expressed $\lambda_{\max}$ in terms of an auxiliary replicated partition function, whose average can be computed in the large-$N$ limit, and derived the distribution $T(u)$ by differentiating with respect to an auxiliary source term, as shown in Eqs.~\eqref{eq:lambda_max_limits} and \eqref{eq:Tu_starting_limits}. For a generic symmetric kernel function $f$, this approach gives general expressions for both quantities in terms of the principal eigenfunction of an associated Fredholm operator. We then specialised this general framework to the quadratic distance kernel. In this case, the Fredholm problem reduces to a set of self-consistent equations for the $d+2$ coefficients $(A,\B,C)$, which fully determine both $\langle \lambda_{\max} \rangle$ and the distribution $T(u)$ of the top eigenvector's components. For specific classes of underlying distributions, namely isotropic distributions and distributions with independent and symmetric components, these equations can be solved analytically, yielding explicit closed-form expressions. For the quadratic distance kernel, these results show that the leading spectral properties depend only on low-order moments of the underlying distribution, up to fourth order. They also clarify the distinct roles played by isotropy and independence. For more general distributions, where analytical solutions of the self-consistent system are not accessible, we solved the equations numerically and used the resulting coefficients to obtain theoretical predictions. We tested these predictions through extensive numerical experiments based on direct diagonalisation of large random matrices with quadratic distance kernel. The comparison between theory and numerical results shows a very good level of agreement across all regimes explored, both for the scaling of $\langle \lambda_{\max} \rangle$ and for the full distribution $T(u)$. Overall, our results provide a general framework to characterise the leading spectral properties of symmetric Euclidean random matrices of the form defined in Eq.~\eqref{eq:matrix_X_generic}, for arbitrary underlying distributions of the vectors. Although closed-form solutions are available only in special cases, the method itself remains fully general, as the relevant quantities can always be obtained by solving the associated self-consistent equations numerically. In addition, in Appendix~\ref{sec:appendix_A} we showed that the coefficients $(A,\B,C)$ admit a natural interpretation in terms of an exact finite-rank decomposition of the spectral problem, which reduces the determination of the non-zero eigenvalues and eigenvectors to a finite-dimensional system. We also discussed in Appendix~\ref{sec:appendix_B} how the replica framework can be extended beyond the i.i.d. setting to correlated disorder of Curie--Weiss type. Finally, in Appendix~\ref{sec:appendix_C} we considered a Gaussian kernel as an example beyond the quadratic distance kernel, showing that the general Fredholm formulation can be used directly even when no finite-dimensional reduction is available. 
Several natural extensions of this work can be considered. First, while we have focused on the average largest eigenvalue, the same replica-based framework could in principle be extended to characterise its full distribution in a large deviation sense \cite{noemi,Fyodorov_topology_triv,Satya_large_dev}. Second, the formalism naturally extends to complex matrices, provided the kernel is chosen so that the resulting Euclidean random matrix is Hermitian. A similar methodology could be employed to investigate other extremal spectral observables, in particular the smallest eigenvalue or the second largest, which are also expected to encode relevant structural information \cite{Susca_second_largest}. It would also be interesting to investigate the Inverse Participation Ratio of the top eigenvector to assess on how many of the random points it is supported, as well as the geometry and topological properties of those points.

\section{Acknowledgments}
P.V. acknowledges support from UKRI FLF Scheme (No. MR/X023028/1). P.V. gratefully acknowledges important exchanges about this problem with Fabio D. Cunden, Pierfrancesco Urbani and Tim Rogers.

\section{Author contributions}
All authors contributed equally to the conception, analysis, and writing of this work.

\appendix

\section{Exact finite rank decomposition for distance matrices}\label{sec:appendix_A}
For the quadratic distance kernel in Eq.~\eqref{eq:quadratic_distance_kernel}, the entries of the matrix $X$ defined in Eq.~\eqref{eq:matrix_X_generic} can be rewritten as

\begin{equation}
X_{ij}
=
\frac{|| \x_i-\x_j||^2}{N}
=
\frac{||\x_i||^2+||\x_j||^2-2\x_i\cdot \x_j}{N}\ .
\end{equation}
Let $\vec r, \vec 1\in\mathbb{R}^N$ be defined by

\begin{equation}
\vec r=(||\x_1||^2,\dots,||\x_N||^2)^{\top},
\qquad
\vec 1=(1,\dots,1)^{\top},
\end{equation}
and let $Y\in\R^{N\times d}$ be the data matrix with entries $Y_{i\ell}=x_{i\ell}$. Then

\begin{equation}
X
=
\frac{1}{N}
\left(
\vec r\,\vec 1^{\top}
+
\vec 1\,\vec r^{\top}
-
2YY^{\top}
\right).
\end{equation}
Since the matrices $Y$ and $YY^{\top}$ have rank at most $d$, while both $\vec r\,\vec 1^{\top}$ and $\vec 1\,\vec r^{\top}$ have rank at most one, it follows that the matrix $X$ has rank at most $d+2$

\begin{equation}
\operatorname{rank}(X)
\le
\operatorname{rank}(\vec r\,\vec 1^{\top})
+
\operatorname{rank}(\vec 1\,\vec r^{\top})
+
\operatorname{rank}(YY^{\top})
\le
d+2\ .
\end{equation}
Applying $X$ to a generic vector $\vec v \in \R^N$ yields for each component $i \in \{1,\dots,N\}$

\begin{equation}
\begin{aligned}
(X \vec v)_i&=\frac{1}{N}\sum_{j=1}^N\left(||\x_i||^2+||\x_j||^2-2\x_i\cdot \x_j\right)v_j\\
&=||\x_i||^2\left(\frac{1}{N}\sum_{j=1}^N v_j\right)-2~\x_i\cdot\left(\frac{1}{N}\sum_{j=1}^N \x_jv_j\right)+\frac{1}{N}\sum_{j=1}^N ||\x_j||^2v_j\\
&=A_N||x_i||^2+\B_N\cdot \x_i+C_N\ ,
\end{aligned}
\end{equation}
where we have defined the coefficients

\begin{equation}\label{eq:ABC_finite_N}
A_N\coloneqq\frac{1}{N}\sum_{j=1}^N v_j\ ,
\qquad
\B_N\coloneqq-\frac{2}{N}\sum_{j=1}^N v_j\x_j\in\mathbb{R}^d\ ,
\qquad
C_N\coloneqq\frac{1}{N}\sum_{j=1}^N ||\x_j||^2v_j\ .
\end{equation}
Therefore, any eigenvector $\vec v$ of $X$, corresponding to a non-zero eigenvalue $\lambda \neq 0$, will have components of the form

\begin{equation}\label{eq:eigenvec_finiteN}
v_i=\frac{A_N||\x_i||^2+\B_N\cdot \x_i+C_N}{\lambda}\ ,
\end{equation}
and $\lambda$ can be obtained recalling that in the main paper we used the normalisation $|| \vec v||^2=\vec v^\top \vec v=N$

\begin{equation}
\begin{aligned}
\lambda \vec v = X\vec v
&\iff \lambda \vec v^{\top}\vec v=\vec v^{\top}X\vec v \\
&\iff \lambda 
=
\frac{1}{N}\sum_{i=1}^N v_i (X\vec v)_i \\
&\iff \lambda 
=
\frac{1}{N}\sum_{i=1}^N v_i\left(A_N||\x_i||^2+\B_N\cdot \x_i+C_N\right) \\
&\iff \lambda 
=
A_N\frac{1}{N}\sum_{i=1}^N ||\x_i||^2v_i
+
\B_N\cdot \frac{1}{N}\sum_{i=1}^N \x_i v_i
+
C_N\frac{1}{N}\sum_{i=1}^N v_i \\
&\iff \lambda 
=
2A_NC_N-\frac{||\B_N||^2}{2}\ .\label{eq:lambda_finite_N}
\end{aligned}
\end{equation}
Thus, the eigenvalue problem for the $N\times N$ matrix $X$ is reduced, for every non-zero eigenvalue, to the determination of the $d+2$ coefficients $A_N$, $\B_N$ and $C_N$. Once these coefficients are known, also the corresponding eigenvector is immediately recovered from Eq.~\eqref{eq:eigenvec_finiteN}. This reduction is a direct consequence of the finite-rank decomposition of $X$. The coefficients $A_N$, $\B_N$ and $C_N$ are determined self-consistently by substituting the eigenvector representation in Eq.~\eqref{eq:eigenvec_finiteN} back into the coefficients definition in Eq.~\eqref{eq:ABC_finite_N}. This gives the finite-$N$ system

\begin{equation}
A_N
=
\frac{1}{\lambda N}
\sum_{i=1}^N
\left(
A_N||\x_i||^2+\B_N\cdot \x_i+C_N
\right)\ ,
\end{equation}
\begin{equation}
\B_N
=
-\frac{2}{\lambda N}
\sum_{i=1}^N
\x_i
\left(
A_N||\x_i||^2+\B_N\cdot \x_i+C_N
\right)\ ,
\end{equation}
\begin{equation}
C_N
=
\frac{1}{\lambda N}
\sum_{i=1}^N
||\x_i||^2
\left(
A_N||\x_i||^2+\B_N\cdot \x_i+C_N
\right)\ ,
\end{equation}
that together with Eq.~\eqref{eq:lambda_finite_N} form a closed system for the $d+2$ unknowns $A_N$, $\B_N$ and $C_N$. In the large-$N$ limit, assuming that the vectors $\x_i$ are i.i.d. with common distribution $P(\x)$, the empirical averages converge to averages with respect to $P(\x)$. The system of equations above, together with the general expression for any eigenvalue $\lambda$ in \eqref{eq:lambda_finite_N}, are fully compatible with Eqs.  \eqref{eq:ABC_sistemone_generale} and \eqref{eq:lambda_final_short} of the main text, once the Perron-Frobenius condition is employed to select the largest positive solution for $\lambda$. Using Eq.~\eqref{eq:eigenvec_finiteN}, the empirical distribution of the components of an eigenvector with non-zero eigenvalue $\lambda$ is given by

\begin{equation}
T_N(u)=\frac{1}{N}\sum_{i=1}^N \delta\left(u-\frac{A_N||\x_i||^2+\B_N\cdot \x_i+C_N}{\lambda}\right)\xrightarrow[N\to\infty]{} \int d\x\,P(\x)\,\delta\left(u-\frac{A||\x||^2+\B\cdot \x+C}{\lambda}\right)\ ,\label{TNuConvergence}
\end{equation}
after the large-$N$ limit is taken for i.i.d. variables. The form in Eq. \eqref{TNuConvergence} is again fully compatible with Eq.~\eqref{eq:T(u)_suitable_for_iid_quadratic_kernel} of the main text for the distribution of top eigenvector's components. This confirms that the limiting distribution of the top eigenvector components is completely determined by the same $d+2$ coefficients that determine the largest eigenvalue.

The same finite-rank structure allows one to rewrite the non-zero spectrum of $X$ in terms of a smaller $(d+2) \times (d+2)$ matrix. Define the feature vectors $\phi(\x)=\left(||\x||^2,x_1,\dots,x_d,1\right)^{\top}\in\R^{d+2}$, and let $\Phi\in\R^{N\times(d+2)}$ be the matrix whose $i$-th row is $\phi(\x_i)^{\top}$. Introducing the $(d+2)\times(d+2)$ matrix

\begin{equation}
M=
\begin{pmatrix}
0 & 0 & 1\\
0 & -2I_d & 0\\
1 & 0 & 0
\end{pmatrix}\ ,
\end{equation}
since $\phi(\x_i)^{\top}M\phi(\x_j)=||\x_i-\x_j||^2$, one has

\begin{equation}
X=\frac{1}{N}\Phi M\Phi^{\top}\ .
\end{equation}
For $\lambda\neq 0$, the Sylvester's determinant identity yields 

\begin{equation}
\begin{aligned}
\det(\lambda I_N-X)
&=\det\left(\lambda I_N-\frac{1}{N}\Phi M\Phi^{\top}\right)\\
&=\lambda^N\det\left(I_N-\frac{1}{\lambda N}\Phi M\Phi^{\top}\right)\\
&=\lambda^N\det\left(I_{d+2}-\frac{1}{\lambda N}M\Phi^{\top}\Phi\right)\\
&=\lambda^{N-(d+2)}\det\left(\lambda I_{d+2}-\frac{1}{N}M\Phi^{\top}\Phi\right).
\end{aligned}
\end{equation}
Therefore, the non-zero eigenvalues of $X$ coincide with the eigenvalues of the $(d+2)\times(d+2)$ matrix $K_N=MG_N$ where

\begin{equation}
K_N\coloneq \frac{1}{N}M\Phi^{\top}\Phi = MG_N\ ,
\qquad
G_N\coloneq \frac{1}{N}\Phi^{\top}\Phi\ .
\end{equation}
Thus, the original $N\times N$ eigenvalue problem is reduced to a $(d+2)\times(d+2)$ eigenvalue problem. Moreover, the corresponding eigenvectors are recovered from the eigenvectors of the reduced matrix. Indeed, if $\underline a\in\R^{d+2}$ satisfies

\begin{equation}
K_N\underline a=\lambda \underline a\ ,\qquad \lambda\neq 0\ ,
\end{equation}
then $\vec v=\Phi\underline a$ satisfies

\begin{equation}
X\Phi\underline a=\frac{1}{N}\Phi M\Phi^{\top}\Phi\underline a=\Phi K_N\underline a=\lambda \Phi\underline a\ .
\end{equation}
Equivalently, every eigenvector of $X$ associated with a non-zero eigenvalue belongs to the column span of $\Phi$, consistently with Eq.~\eqref{eq:eigenvec_finiteN}. As a final remark, the matrix $G_N$ is the empirical moment matrix of the features. Explicitly

\begin{equation}
G_N=\frac{1}{N}\sum_{i=1}^N \phi(\x_i)\phi(\x_i)^{\top}\ .
\end{equation}
Since $\phi(\x)$ contains the entries $||\x||^2$, $x_1,\dots,x_d$ and $1$, the entries of $G_N$ involve only empirical moments of the coordinates of $\x$ up to order four. If the vectors $\x_i$ are i.i.d. with common distribution $P(\x)$, then, in the large-$N$ limit

\begin{equation}
G_N \xrightarrow[N\to\infty]{} G=\left\langle \phi(\x)\phi(\x)^{\top}\right\rangle\ ,
\end{equation}
and hence

\begin{equation}
K_N \xrightarrow[N\to\infty]{} K=MG\ .
\end{equation}
Therefore, this confirms that the limiting coefficients $A$, $\B$ and $C$, and in particular the largest eigenvalue, are determined only by the moments of $P(\x)$ up to fourth order.

\section{Beyond i.i.d. assumption, correlated disorder}\label{sec:appendix_B}
In the main text, we considered symmetric Euclidean random matrices constructed from vectors $\{\x\}_{i=1}^N$ independently drawn from a common distribution $P(\x)$. In this Appendix, we extend the analysis to the case of correlated points, focusing on a setting in which the joint distribution of the vectors is of Curie–Weiss type

\begin{equation}
\begin{aligned}\label{eq:P_cure_weiss}
    P(\x_1,\dots,\x_N) = \frac{1}{\mathcal{Z}_N} \exp\left\{-\sum_{i=1}^NV(\x_i) + \frac{J}{2N} \left\| \sum_{i=1}^N \bm\xi(\x_i)\right\|^2 \right\}\ ,
\end{aligned}
\end{equation}
where $\mathcal{Z}_N$ is the normalisation constant, $V:\mathbb{R}^d\to\mathbb{R}$ is a scalar potential, $\bm\xi:\mathbb{R}^d\to\mathbb{R}^p$ is the vector-valued observable coupled by the mean-field interaction, and $J$ is the coupling constant. In particular, we show how the replica framework can be adapted to this setting and discuss the resulting modifications to the largest eigenvalue statistics.  The derivation presented in Section~\ref{sec:eigenvalue} of the main text carries over to this correlated setting up to Eq.~\eqref{eq:Z^n_for_T(u)}. At this stage, the only modification arises in the averaging over the joint distribution, and the averaged replicated partition function is given by

\begin{equation}
\begin{aligned}\label{eq:Z_replicated_curie_weiss}
    \langle [Z_N(\beta)]^n \rangle \propto \int \dd\boldsymbol{\lambda}&\{\mathcal{D}\rho_c\}\{\mathcal{D}\hat \rho_c\} \exp\left\{\ii \frac{\beta}{2}N\sum_{c=1}^n\lambda_c+\frac{\beta}{2}N\sum_{c=1}^n\int \dd\x \dd\y~\rho_c(\x)\rho_c(\y)f(\x,\y)+\ii N\sum_{c=1}^n\int \dd\x~\rho_c(\x) \hat \rho_c(\x)\right\} \times \\
    & \times\int \prod_{c=1}^{n} \dd\Vec{v}_c \prod_{i=1}^N\dd\x_i~P(\x_1,\dots,\x_N) \exp\left\{-\ii\frac{\beta}{2}\sum_{c=1}^n\sum_{i=1}^N \lambda_c v_{ic}^2-\ii\sum_{c=1}^n\sum_{i=1}^N \hat \rho_c(\x_i)v_{ic}\right\} \ .
\end{aligned}
\end{equation}
Introducing the mean-field order parameter

\begin{equation}
\begin{aligned}
    \Om  \coloneqq \frac{1}{N}\sum_{i=1}^N\bm\xi(\x_i) \in \R^p\ ,
\end{aligned}
\end{equation}
and enforcing its definition in Eq. \eqref{eq:Z_replicated_curie_weiss} via functional deltas as

\begin{equation}
\begin{aligned}
    1= \int \dd\Om \dd\hOm   \exp\left\{-N \Om \cdot \hOm + \hOm \cdot\sum_{i=1}^N \bm\xi(\x_i)\right\}\ ,
\end{aligned}
\end{equation}
we can rewrite the integral in the last line of Eq.~\eqref{eq:Z_replicated_curie_weiss} as

\begin{equation}
\begin{aligned}
    I_N &\coloneqq \int \prod_{c=1}^{n} \dd\Vec{v}_c \prod_{i=1}^N\dd\x_i~P(\x_1,\dots,\x_N) ~ \exp\left\{-\ii\frac{\beta}{2}\sum_{c=1}^n\sum_{i=1}^N \lambda_c v_{ic}^2-\ii\sum_{c=1}^n\sum_{i=1}^N \hat \rho_c(\x_i)v_{ic}\right\}\\
    &=\frac{1}{\mathcal{Z}_N} \int \dd\Om \dd\hOm \exp\left\{\frac{NJ}{2} ||\Om||^2-N\Om \cdot \hOm+N \log\left[\int \dd \x ~ \ee^{-V(\underaccent{\bar}{x})+\hOm \cdot \bm\xi(\underaccent{\bar}{x})} \prod_{c=1}^n \int \dd v_c ~ \ee^{- \ii \frac{\beta}{2} \lambda_c v_c^2-\ii \hat\rho_c(\underaccent{\bar}{x})v_c} \right] \right\}\ .
\end{aligned}
\end{equation}
Therefore the averaged replicated partition function can be written in a form amenable to a saddle-point evaluation for large $N$, which, assuming a replica symmetric structure, takes the form

\begin{equation}
\begin{aligned}
    \langle [Z_N(\beta)]^n \rangle \propto  \frac{1}{\mathcal{Z}_N}  \int &\dd\lambda \mathcal{D}\rho \mathcal{D}\hat\rho \dd\Om \dd\hOm~ \exp\left\{ N \mathcal{S}[\lambda,\rho,\hat \rho,\Om,\hOm]\right\}\ ,
\end{aligned}
\end{equation}
where the action is defined as $\mathcal{S}[\lambda,\rho, \hat \rho,\Om,\hOm]=\mathcal{S}_1[\lambda]+\mathcal{S}_2[\rho]+\mathcal{S}_3[\rho,\hat \rho]+\mathcal{S}_4[\lambda,\hat \rho,\Om,\hOm]$, with $\mathcal{S}_1,\mathcal{S}_2$ and $\mathcal{S}_3$ given by Eqs. \eqref{eq:S_1_on}-\eqref{eq:S_3_on} of the main text, and

\begin{equation}
\begin{aligned}\label{eq:S_4_curie_weiss}
   \mathcal{S}_4[\lambda,\hat \rho,\Om,\hOm] = \frac{J}{2} ||\Om||^2-\Om \cdot \hOm+ \log\left[~\int \dd \x ~ \ee^{-V(\underaccent{\bar}{x})+\hOm \cdot \bm\xi(\underaccent{\bar}{x})} \left( \int \dd v ~ \ee^{- \ii \frac{\beta}{2} \lambda v^2-\ii \hat\rho(\underaccent{\bar}{x})v} \right)^n ~ \right]\ .
\end{aligned}
\end{equation}
Defining the auxiliary functions

\begin{equation}
\begin{aligned}
   G(\x) \coloneqq \int \dd v \exp\left\{-\ii \frac{\beta}{2}\lambda v^2-\ii \hrho(\x)v\right\}\ ,
\end{aligned}
\end{equation}

\begin{equation}
\begin{aligned}
  Z_n(\hOm) &\coloneqq \int \dd \x ~ \exp\left\{-V(\underaccent{\bar}{x})+\hOm \cdot \bm\xi(\underaccent{\bar}{x})\right\} G(\x)^n\\
  &= Z_0(\hOm)+n \int \dd \x ~ \exp\left\{-V(\underaccent{\bar}{x})+\hOm \cdot \bm\xi(\underaccent{\bar}{x})\right\} \log[G(\x)] + o(n)\ ,
\end{aligned}
\end{equation}

\begin{equation}
\begin{aligned}
  S_n(\Om,\hOm) &\coloneqq \frac{J}{2} ||\Om||^2-\Om \cdot \hOm + \log(Z_n(\hOm))\\
  &=S_0(\Om,\hOm)+n \int \dd \x~ P_{\hOm}(\x) \log[G(\x)]+o(n)\ ,
\end{aligned}
\end{equation}
and the tilted single-site distribution

\begin{equation}
\begin{aligned}\label{eq:P_omega_curie_weiss}
  P_{\hOm}(\x) \coloneqq \frac{1}{Z_0(\hOm)}\exp\{-V(\x)+\hOm \cdot \bm\xi(\x)\}=\frac{\exp\{-V(\x)+\hOm \cdot \bm\xi(\x)\}}{\int \dd \y~\exp\{-V(\y)+\hOm \cdot \bm\xi(\y)\}}\ ,
\end{aligned}
\end{equation}
we can rewrite the normalisation constant of the joint distribution in Eq.~\eqref{eq:P_cure_weiss} as

\begin{equation}
\begin{aligned}\label{eq:termine_1}
  \mathcal{Z}_N= \int \dd\Om \dd\hOm~\exp\{NS_0(\Om,\hOm)\}\ ,
\end{aligned}
\end{equation}
and expand the functional $\mathcal{S}_4[\lambda,\hrho,\Om,\hOm]$ in Eq.~\eqref{eq:S_4_curie_weiss} to the first order in $n$ 

\begin{equation}
\begin{aligned}\label{eq:termine_2}
   \mathcal{S}_4[\lambda,\hat \rho,\Om,\hOm] = S_0(\Om,\hOm)+n \int \dd \x~ P_{\hOm}(\x) \log[G(\x)]+o(n)\ .
\end{aligned}
\end{equation}
Eqs.~\eqref{eq:termine_1} and \eqref{eq:termine_2} show that, when computing $\langle \lambda_{\mathrm{max}} \rangle$ via Eq.~\eqref{eq:lambda_max_limits}, the contribution of $S_0(\Om,\hOm)$ inside $\mathcal{S}_4[\lambda,\hat \rho,\Om,\hOm]$ is exactly canceled by the normalization constant $\mathcal{Z}_N$ after the saddle-point evaluation. As a result, the two terms do not contribute to the average largest eigenvalue. The saddle-point equations for the mean-field order parameter and its conjugate field read 

\begin{equation}
\begin{aligned}\label{eq:saddle_omega_curie_weiss}
   0= \frac{\partial\mathcal{S}}{\partial \Om}=J\Om-\hOm \iff \hOm=J\Om\ ,
\end{aligned}
\end{equation}

\begin{equation}
\begin{aligned}\label{eq:saddle_hat_omega_curie_weiss}
   0= \frac{\partial\mathcal{S}}{\partial \hOm}= -\Om+\frac{1}{Z_0(\hOm)}\frac{\partial Z_0(\hOm)}{\partial \hOm}+\mathcal{O}(n) \iff \Om=\int \dd\x ~ P_{\hOm}(\x) \bm\xi(\x)=\langle \bm\xi(\x)\rangle_{\hOm}\ ,
\end{aligned}
\end{equation}
where $\langle ~ \cdot ~ \rangle_{\hOm}$ denotes the average with respect to the tilted single-site distribution $P_{\hOm}$ defined in Eq.~\eqref{eq:P_omega_curie_weiss}. In deriving the saddle-point equation with respect to the conjugate field $\hOm$, we have neglected the term linear in $n$ appearing in $\mathcal{S}_4$ in Eq.~\eqref{eq:termine_2}. This is justified since, in the computation of $\langle \lambda_{\mathrm{max}} \rangle$ via Eq.~\eqref{eq:lambda_max_limits}, such contributions vanish in the limit $n \to 0$. For the same reason, this term does not affect the saddle-point equations for the remaining variables. Combining Eqs.~\eqref{eq:saddle_omega_curie_weiss} and \eqref{eq:saddle_hat_omega_curie_weiss} yields a closed self-consistency equation for the mean-field order parameter

\begin{equation}
\begin{aligned}\label{eq:Omega_self_consist}
   \Om=\langle \bm\xi(\x)\rangle_{J\Om} = \frac{\int \dd\x ~ \exp\{-V(\x)+J\Om \cdot \bm\xi(\x)\} ~\bm\xi(\x)}{\int \dd \x~\exp\{-V(\x)+J\Om \cdot \bm\xi(\x)\}}\ .
\end{aligned}
\end{equation}
We now observe that the second term of $\mathcal{S}_4[\lambda,\hat \rho,\Om,\hOm]$ in Eq.~\eqref{eq:termine_2} is identical to $\mathcal{S}_4[\lambda,\hat \rho]$ in Eq.~\eqref{eq:S_4_on} of the main text, with the sole replacement of the original single-site distribution $P(\x)$ by the tilted single-site distribution $P_{J\Om}(\x)$ defined in Eq.~\eqref{eq:P_omega_curie_weiss}. Moreover, the remaining contributions $\mathcal{S}_1$, $\mathcal{S}_2$, and $\mathcal{S}_3$ are unchanged with respect to the i.i.d. calculation of Section~\ref{sec:eigenvalue}. It follows that the saddle-point equations for $\lambda$, $\rho$, and $\hrho$, together with their solutions, are formally the same as those derived in Eqs.~\eqref{eq:saddle_point_rho}--\eqref{eq:fine_curie_weiss}, upon replacing $P(\x)$ by $P_{J\Om}(\x)$ throughout. In particular, Eq.~\eqref{eq:Fredholm_problem} still holds, provided the operator $T_f$ in Eq.~\eqref{eq:operator_Tf_iid} is replaced by 
\begin{equation}
\begin{aligned}
    (T_{f,J\Om}\psi)(\x)\coloneq \int \dd\y~P_{J\Om}(\y)f(\x,\y)\psi(\y)\ .
\end{aligned}
\end{equation}
Consequently, in the quadratic distance kernel case, the expression for the largest eigenvalue in Eq.~\eqref{eq:lambda_final_short} remains unchanged, namely

\begin{equation}
\begin{aligned}\label{eq:lambda_max_curie_weiss}
    \langle \lambda_{\mathrm{max}}\rangle=2AC-\sum_{\ell=1}^d \frac{B^2_{\ell}}{2}\ ,
\end{aligned}
\end{equation}
and the coefficients $(A,\B,C)$ are determined by the same self-consistency equations as in Eq.~\eqref{eq:ABC_sistemone_generale} of the i.i.d. case, with all averages now taken with respect to the tilted measure $P_{J\Om}$. Explicitly

\begin{equation}
\begin{cases}\label{eq:ABC_sistemone_curie_weiss}
\displaystyle
A = \frac{\sum_{\ell} B_\ell \langle x_\ell \rangle_{J\Om} + C}
{\sqrt{s} - \sum_{\ell} \langle x_\ell^2 \rangle_{J\Om}}\ , \\[1.2em]

\displaystyle
B_q = -2 \,
\frac{
A \sum_{\ell} \langle x_\ell^2 x_q \rangle_{J\Om}
+ \sum_{\ell \ne q} B_\ell \langle x_\ell x_q \rangle_{J\Om}
+ C \langle x_q \rangle_{J\Om}
}{
\sqrt{s} + 2 \langle x_q^2 \rangle_{J\Om}
}\ , \quad \forall q=1,\dots,d\ , \\[1.2em]

\displaystyle
C =
\frac{
A \sum_{\ell,q} \langle x_\ell^2 x_q^2 \rangle_{J\Om}
+ \sum_{q} B_q \sum_{\ell} \langle x_\ell^2 x_q \rangle_{J\Om}
}{
\sqrt{s} - \sum_{\ell} \langle x_\ell^2 \rangle_{J\Om}
}\ .
\end{cases}
\end{equation}
Thus, after determining the mean-field order parameter $\Om$ self-consistently from Eq.~\eqref{eq:Omega_self_consist}, the computation of $\langle \lambda_{\mathrm{max}}\rangle$ proceeds exactly as in the i.i.d. case: one solves the same finite-dimensional system for the coefficients $(A,\B,C)$, with all averages taken with respect to $P_{J\Om}$, and then substitutes the resulting coefficients into Eq.~\eqref{eq:lambda_max_curie_weiss}.

\section{Gaussian kernel}\label{sec:appendix_C}
\begin{figure}[htbp]
    \centering
    \includegraphics[width=0.6\textwidth]{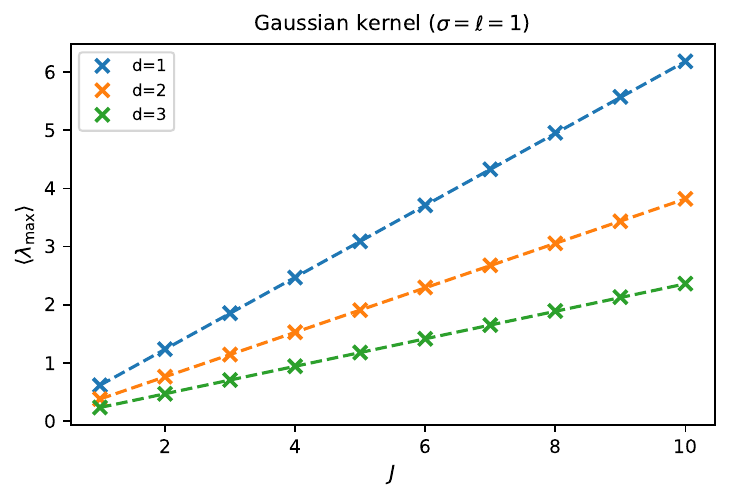}
    \caption{Scaling of $\langle \lambda_{\max} \rangle$ for the Gaussian kernel in Eq.~\eqref{eq:gaussian_centered_kernel}, with vectors drawn from the Gaussian distribution in Eq.~\eqref{eq:gaussian_P_appendixC}. Comparison between the theoretical prediction given by Eq.~\eqref{eq:appendix_C_lambda_max} (dashed lines) and the values obtained via direct numerical diagonalisation (colored crosses), as a function of $J$ for different dimensions $d$ (colors), with $\sigma=\ell=1$. Numerical results are obtained for matrices of size $N=1500$ and averaged over $100$ independent realisations.}    
    \label{fig:gaussian_kernel_lambda_max}
\end{figure}

In the main text we focused on the quadratic distance kernel, for which the general Fredholm equation reduces to the finite-dimensional system in Eq.~\eqref{eq:ABC_sistemone_generale} for the $d+2$ coefficients $(A,\B,C)$. This reduction makes it possible to obtain explicit predictions for the largest eigenvalue and for the distribution of the top eigenvector's components. In this Appendix, we consider a different kernel choice, showing how the general Fredholm formulation developed in the main text can be used beyond the quadratic distance kernel. In this case no finite-dimensional reduction occurs, and the relevant quantities are obtained directly from the eigenvalue problem in Eq.~\eqref{eq:Fredholm_problem} for the integral operator $T_f$. Specifically, we study the symmetric Euclidean random matrix in Eq.~\eqref{eq:matrix_X_generic} with Gaussian kernel
\begin{equation}
\begin{aligned}\label{eq:gaussian_centered_kernel}
    f(\x,\y)=J\exp\left\{-\frac{\|\x-\y\|^2}{2\ell^2}\right\}\ , \qquad J>0\ .
\end{aligned}
\end{equation}
We assume that the vectors $\{\x_i\}_{i=1}^N$ are drawn i.i.d.\ from the centered Gaussian distribution
\begin{equation}
\begin{aligned}\label{eq:gaussian_P_appendixC}
    P(\x)=\frac{1}{(2\pi\sigma^2)^{d/2}}\exp\left\{-\frac{\|\x\|^2}{2\sigma^2}\right\}\ .
\end{aligned}
\end{equation}
With these choices, the Fredholm equation~\eqref{eq:Fredholm_problem} becomes
\begin{equation}
\begin{aligned}\label{eq:Fredholm_gaussian_kernel}
\sqrt{s} \psi(\x)
&= \int \dd\y~P(\y)f(\x,\y)\psi(\y)= \frac{J}{(2\pi\sigma^2)^{d/2}}\int \dd\y~
\exp\left\{-\frac{||\y||^2}{2\sigma^2}-\frac{||\x-\y||^2}{2\ell^2}\right\}\psi(\y)\ .
\end{aligned}
\end{equation}
It is clear that this integral equation is solved by the Gaussian ansatz
\begin{equation}
\begin{aligned}\label{eq:gaussian_ansatz_psi}
\psi(\x)=C\exp\left\{-\frac{a||\x||^2}{2}\right\}\ ,
\end{aligned}
\end{equation}
where $C$ is a normalization constant and $a>0$ is to be determined. For the Gaussian kernel in Eq.~\eqref{eq:gaussian_centered_kernel}, the
positivity-improving condition in Eq.~\eqref{eq:positivity_improving} readily holds. Therefore, as
discussed in the main text, Jentzsch's theorem applies: the compact
positivity-improving integral operator $T_f$ has a strictly positive and
simple principal eigenvalue, and its associated eigenfunction is the unique
strictly positive eigenfunction, up to multiplication by a positive constant.
Since the Gaussian ansatz in Eq.~\eqref{eq:gaussian_ansatz_psi} is strictly
positive and solves the Fredholm equation exactly, it cannot correspond to a
subleading eigenvalue. It therefore identifies the principal eigenfunction,
and the associated eigenvalue is automatically the desired average largest
eigenvalue, $\langle\lambda_{\max}\rangle=\sqrt{s}$. Inserting Eq.~\eqref{eq:gaussian_ansatz_psi} into Eq.~\eqref{eq:Fredholm_gaussian_kernel} gives
\begin{equation}
\begin{aligned}\label{eq:gaussian_kernel_self_const}
\sqrt{s}\exp\left\{-\frac{a||\x||^2}{2}\right\}&=
\frac{J}{(2\pi\sigma^2)^{d/2}}\int \dd\y~
\exp\left\{-\frac{||\y||^2}{2\sigma^2}
-\frac{||\x-\y||^2}{2\ell^2}
-\frac{a||\y||^2}{2}\right\} \\
&= \frac{J}{(\sigma^2 q)^{d/2}}
    \exp\left\{-\left(\frac{1}{\ell^2}-\frac{1}{q\ell^4}\right)\frac{\|\x\|^2}{2}\right\}\ ,
\end{aligned}
\end{equation}
where we have performed the Gaussian $\y$-integral and introduced the auxiliary parameter
\begin{equation}
\begin{aligned}\label{eq:q_appendix_C}
q \coloneqq a+\frac{1}{\ell^2}+\frac{1}{\sigma^2}= \frac{\ell^2+\sigma^2+\ell^2\sigma^2a}{\ell^2\sigma^2}\ .
\end{aligned}
\end{equation}
For Eq.~\eqref{eq:gaussian_kernel_self_const} to hold for every $\x\in\mathbb{R}^d$, the coefficients of $\|\x\|^2$ in the two exponentials must coincide
\begin{equation}
\begin{aligned}\label{eq:a_appendix_C}
    a=\frac{1}{\ell^2}-\frac{1}{q\ell^4}
    \iff 
    a=\frac{1}{2\sigma^2}\left[\sqrt{1+\frac{4\sigma^2}{\ell^2}}-1\right]\ ,
\end{aligned}
\end{equation}
while matching the prefactors fixes
\begin{equation}
\begin{aligned}
    \sqrt{s}=\frac{J}{(\sigma^2 q)^{d/2}}\ .
\end{aligned}
\end{equation}
Combining Eqs.~\eqref{eq:q_appendix_C} and \eqref{eq:a_appendix_C} yields
\begin{equation}
\begin{aligned}
    \sigma^2 q
    =\frac{1}{4}\left(1+\sqrt{1+\frac{4\sigma^2}{\ell^2}}\right)^2\ ,
\end{aligned}
\end{equation}
and therefore, from Eq.~\eqref{eq:lambda_max_final_long} of the main text, we obtain the average largest eigenvalue
\begin{equation}
\begin{aligned}\label{eq:appendix_C_lambda_max}
    \langle \lambda_{\mathrm{max}}\rangle
    =\sqrt{s}
    =J\left(\frac{2}{1+\sqrt{1+\frac{4\sigma^2}{\ell^2}}}\right)^d\ .
\end{aligned}
\end{equation}
The normalization constant $C$ in Eq.~\eqref{eq:gaussian_ansatz_psi} is fixed by the definition of $s$ in Eq.~\eqref{eq:psi_def_from_rho} of the main text, which directly implies
\begin{equation}
\begin{aligned}
    1=\int \dd\x~P(\x)\left(\frac{\psi(\x)}{\sqrt{s}}\right)^2\ .
\end{aligned}
\end{equation}
Using Eqs. \eqref{eq:gaussian_P_appendixC} and \eqref{eq:gaussian_ansatz_psi}, we find
\begin{equation}
\begin{aligned}
    1
    &=\frac{C^2}{s(2\pi\sigma^2)^{d/2}}\int \dd\x~\exp\left\{-\left(a+\frac{1}{2\sigma^2}\right)\|\x\|^2\right\} \\
    &=\frac{C^2}{s}\left(1+2a\sigma^2\right)^{-d/2}\ ,
\end{aligned}
\end{equation}
and therefore
\begin{equation}
\begin{aligned}
    C=\sqrt{s}\left(1+2a\sigma^2\right)^{d/4}
    =\sqrt{s}\left(1+\frac{4\sigma^2}{\ell^2}\right)^{d/8}\ ,
\end{aligned}
\end{equation}
where in the last equality we used Eq.~\eqref{eq:a_appendix_C}. The density of the top eigenvector's components is then obtained from the general formula in Eq.~\eqref{eq:T(u)_suitable_for_iid} of the main text, namely
\begin{equation}
\begin{aligned}
    T(u)=\int \dd\x~P(\x)\delta\left(u-\frac{\psi(\x)}{\sqrt{s}}\right)\ ,
\end{aligned}
\end{equation}
with $\psi(\x)$ given by Eq.~\eqref{eq:gaussian_ansatz_psi} and the constants fixed as above. In Fig.~\ref{fig:gaussian_kernel_lambda_max} we compare the theoretical prediction for $\langle \lambda_{\max} \rangle$, given by Eq.~\eqref{eq:appendix_C_lambda_max} (dashed lines), with the values obtained via direct numerical diagonalisation (coloured crosses), as a function of $J$ and for different dimensions $d=1,2,3$ depicted with different colours. The numerical results are obtained from matrices of size $N=1500$, averaging over $100$ independent realisations for each value of $J$, with $\sigma=\ell=1$, and show very good agreement with the theoretical prediction.

\bibliography{biblio}

\end{document}